\documentclass[pra,twocolumn,nofootinbib,floatfix,10pt]{revtex4-1}
\pdfoutput=1 
\usepackage[utf8]{inputenc}
\usepackage[T1]{fontenc}

\usepackage{graphicx}
\usepackage{amsmath, amsxtra, amssymb, mathtools, isomath, nccmath, xspace}
\usepackage{ wasysym }
\usepackage{setspace, changepage}
\usepackage{bbold}
\usepackage{hhline}
\usepackage{multirow}
\usepackage{hyperref}
\usepackage[leftcaption]{sidecap}

\newcommand{\ket}[1]{\ensuremath{\left| #1 \right\rangle}\xspace}

\long\def\symbolfootnote[#1]#2{\begingroup%
\def\thefootnote{\fnsymbol{footnote}}\footnotetext[#1]{#2}\endgroup}

\usepackage{epsfig}

\date{\today}

\begin{document}




\title{  
Revealing Hidden Antiferromagnetic Correlations in Doped Hubbard Chains via String Correlators
}

\author{Timon~A.~Hilker$^{1 \dag}$}%
\author{Guillaume~Salomon$^{1}$}%
\author{ Fabian~Grusdt$^{2}$}%
\author{Ahmed~Omran$^{1}$}%
\author{ Martin~Boll$^{1}$}%
\author{ Eugene~Demler$^{2}$}%
\author{Immanuel~Bloch$^{1,3}$}%
\author{Christian~Gross$^{1}$}%

\affiliation{$^{1}$Max-Planck-Institut f\"{u}r Quantenoptik, 85748 Garching, Germany}
\affiliation{$^{2}$Department of Physics, Harvard University, Cambridge, MA 02138,USA}
\affiliation{$^{3}$Fakult\"{a}t f\"{u}r Physik, Ludwig-Maximilians-Universit\"{a}t, 80799 M\"{u}nchen, Germany}

\symbolfootnote[2]{Electronic address: {timon.hilker@mpq.mpg.de}}







\begin{abstract}
Topological phases, like the celebrated Haldane phase in spin-1 chains,
defy characterization through local order parameters. Instead, non-local string
order parameters can be employed to reveal their hidden order. Similar diluted
magnetic correlations appear in doped one-dimensional lattice systems due to the
phenomenon of spin-charge separation. Here we report on the direct observation
of such hidden magnetic correlations via quantum gas microscopy of hole-doped
ultracold Fermi-Hubbard chains. The measurement of non-local spin-density
correlation functions reveals a hidden finite-range antiferromagnetic order, a
direct consequence of spin-charge separation. Our technique demonstrates how
topological order can directly be measured in experiments and it can be
extended to higher dimensions to study the complex interplay between
magnetic order and density fluctuations.
\end{abstract}
\maketitle

The Fermi-Hubbard model, describing systems of strongly correlated fermions on a
lattice, lies at the heart of our understanding of the Mott insulator-metal
transitions and quantum magnetism~\cite{auerbach1994}. The complexity of the
interplay between hole doping and magnetic ordering in this model is believed to
give rise to a rich phase diagram, including a High-Tc superconducting phase as,
for example, observed in cuprate compounds \cite{Lee2006}. In one dimension
however, the competition between the spin and density sectors is largely absent
due to the separation of the spin and density modes at low energy. This
phenomenon of spin-charge separation, generally appearing in Luttinger liquids,
is well understood theoretically~\cite{giamarchi2004}, but there are only
limited experimental observations. All experimental evidences of this
foundational phenomenon are based so far on spectroscopic~\cite{kim1996,
segovia1999,kim2006} or transport measurements~\cite{auslaender2005, jompol2009}
in condensed matter systems. Nevertheless, quasi long-range antiferromagnetic
order at zero temperature, as conventionally measured by two-point spin
correlation functions, gets suppressed by a finite hole density in the system.
However, due to the independence of the spin and charge sectors the order is not
truly reduced, but rather hidden~\cite{denNijs1989,kennedy1992b,kruis2004a}. It
can be revealed by measurements over an extensive part of the system
allowing to construct string correlation functions. In analogy to the spin-1
Haldane phase~\cite{haldane1983,haldane1983b, affleck1987} this requires measuring all spins in the
chain. A closely related way to unveil the hidden order is to work directly in
"squeezed space", where empty sites are completely removed from the
system~\cite{ogata1990,ren1993,zaanen2001,kruis2004}. In traditional condensed
matter systems neither string order can be measured, nor is squeezed space
accessible to experiments. Fermionic quantum gas
microscopes~\cite{haller2015,cheuk2016,parsons2016,omran2015,edge2015,brown2016xx}, in contrast, give access to snapshots of
the full spin and density distribution with single site
resolution~\cite{boll2016}, such that non-local correlation functions can be
extracted~\cite{endres2011}. Here we report on the direct measurement of string
correlations in ultracold Fermi-Hubbard chains. The ability to locally detect
holes, doublons and the spin state allows for an analysis of the system directly
in squeezed space, in which Heisenberg spin correlations are
restored\cite{kruis2004}. Our observations provide a microscopic picture of
spin-charge separation independent of the more frequently discussed spectral
properties or the excitation dynamics.

\begin{figure*}[t!]
\centering
\includegraphics[]{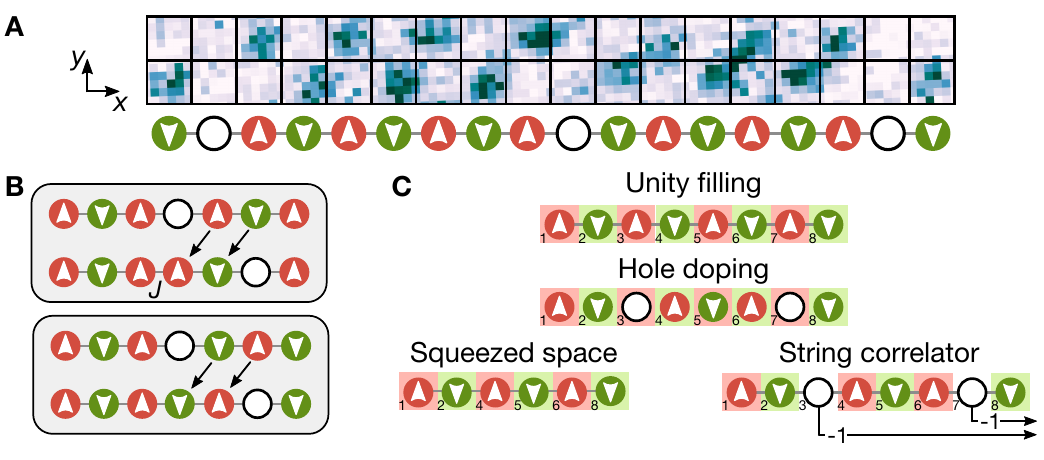}

\caption{
\textbf{Analysis of a doped Hubbard chain.}
\textbf{(A)} Experimental spin and density resolved picture of a single,
slightly-doped Hubbard chain after a local Stern-Gerlach-like detection. The
reconstructed chain is shown below the picture. \textbf{(B)} Illustration of
the magnetic environment around a hole. For aligned spins the hole cannot
freely delocalize due to the magnetic energy cost $J$, which is absent for
anti-aligned spins. \textbf{(C)} Illustration of hole induced AFM parity flips,
squeezed space and string correlator. Hole doping leads to AFM parity flips
highlighted by the color mismatch between the spins and the background (top).
Squeezed space is constructed by removing all sites with holes from the chain
(bottom left). In the string correlator analysis the flip in the AFM parity is
canceled by a multiplication of $-1$ for each hole (bottom right). Comparing
either of these analyses to the conventional two-point correlator reveals the
hidden finite-range AFM order in the system.
}
\end{figure*}
We probed the physics of the doped one-dimensional Fermi-Hubbard model using a
balanced spin mixture of $^{6}$Li trapped in a single plane of a two-dimensional
optical lattice. A versatile quantum gas microscope allowed for the
simultaneous local detection of both spin states~\cite{boll2016}~(see Fig.~1A).
By controlling the lattices depths in the different spatial
directions~\cite{som} we created independent one-dimensional systems described by the single
band Hubbard Hamiltonian
\begin{equation*} \hat{H} = -t \sum_{i,\sigma} (\hat{c}^\dagger_{i,\sigma}
\hat{c}_{i+1,\sigma} + \textrm{h.c.}) + U \sum_i
\hat{n}_{i,\uparrow}\hat{n}_{i,\downarrow} + \sum_{i,\sigma} \epsilon_{i}
\hat{n}_{i,\sigma}\, \end{equation*} 
The fermion creation (annihilation) operator is denoted by
$\hat{c}^\dagger_{i,\sigma}$ ($\hat{c}_{i,\sigma}$) at site $i$ for each of the
two spin states $\sigma =\ \uparrow, \downarrow$ and the operator
$\hat{n}_{i,\sigma}=\hat{c}^\dagger_{i,\sigma}\hat{c}_{i,\sigma}$ counts the
number of atoms with spin $\sigma$ at site $i$. The energy offsets $\epsilon_i$
result from an additional confinement due to the lattice beams, which leads to
a smoothly changing local density ${n_i=\langle \hat{n}_i \rangle =
\langle\hat{n}_{i,\uparrow}+\hat{n}_{i,\downarrow}\rangle}$. At half filling in
the strong coupling limit ($U/t \gg 1$) the Fermi-Hubbard model reduces to a
Heisenberg spin chain with $J=4t^2/U$ and supports quasi long-range antiferromagnetic (AFM)
order at zero temperature~\cite{auerbach1994}. The doped system is described at
long wavelength by Luttinger liquid theory, which predicts at zero temperature
an algebraic decay of the spin correlations with distance that is faster than
the one of the Heisenberg model~\cite{giamarchi2004}. This decay can be understood from
spin-charge separation, allowing  holes to freely move in the AFM spin-chain.
Consequently, the spins around the hole are anti-aligned and the sign of the
staggered magnetization $(-1)^iS_i^z$, called AFM parity, changes. This implies
that a hole acts as a domain wall of the AFM parity, which reduces the spin
correlations. The spin order however, is still present and can be revealed
either in squeezed space by effectively removing the holes in the analysis or
by evaluating string correlators, which take the AFM parity domain walls into
account by flipping the sign of the correlator (Fig.~1C). Analytic and
numerical studies\cite{kruis2004} have shown that at zero temperature, the
two-point spin correlations in squeezed space are comparable to the ones of a pure Heisenberg chain, for any doping and any repulsive interaction $U$. This is readily understood in the $U/t\rightarrow\infty$ 
limit, where the many-body wave function $\Psi(\{x_{j,\sigma}\}) =
\Psi_{\rm ch}(\{x_j\})\,\Psi_{\rm s}(\{\tilde{x}_{j,\sigma}\})$ factorizes exactly into a
density $\Psi_{\rm ch}$ and a spin $\Psi_{\rm s}$ part~\cite{woynarovich1982,ogata1990}.
The spin degree of freedom is described by a Heisenberg model in squeezed
space with the spins "living" on a lattice defined by the positions of spinless, non-interacting
fermions~\cite{ren1993}. 
Distances in squeezed space are rescaled by the spinless
fermion density $\tilde{x} \sim n x$. Also at non-zero temperature and finite interactions, the spin
correlations in squeezed space are governed by a Heisenberg model with a
renormalized exchange coupling $J_{\textrm{eff}}(n)$ that depends on the
original density $n$~\cite{som}.

The experiment started with a two-dimensional degenerate two-component
Fermi gas. Using the large spacing component of an optical superlattice  ($a_{sl} = 2.3\,\mu$m), the system was divided into about ten independent one-dimensional tubes. The Fermi Hubbard chains were then realized using a lattice
of $1.15\,\mu$m spacing along the tubes. The atom number was set such that the maximum
density in the chains was typically just below unity. At the final lattices depths the
tunneling amplitude reached $t = h \times 400\,$Hz, and the confinement due to the
lattice beams fixed the length of the central tubes to about $15$ sites. The
onsite repulsion $U$ was tuned to $h \times 2.9\,$kHz using the broad Feshbach
resonance between the hyperfine states $\ket{\downarrow} = \ket{F, m_F}=
\ket{1/2,-1/2}$ and $\ket{\uparrow} = \ket{1/2, 1/2}$ to set a scattering
length of $2000$ Bohr radii at the end of the lattice ramps. These parameters
and the lattice ramps have been optimized to produce cold, strongly interacting
doped Hubbard chains~\cite{som}. For the detection of the spin and density
degrees of freedom the lattice depth along the tubes was rapidly increased, followed by a local Stern-Gerlach like detection
using a magnetic field gradient and the short scale component of a superlattice
transverse to the tubes~\cite{boll2016}. Applying Raman sideband cooling for
$500\,$ ms, we collected fluorescence photons on an EMCCD camera to form a high
contrast and site resolved image of the atomic distribution~\cite{omran2015} as
shown in Fig.~1A. From comparison of the measured spin-correlations at half
filling to Quantum Monte-Carlo results~\cite{boll2016}, we estimated the
temperature in the central chains to be $0.51(2)\,t$ or $0.90(3)\,J$, which
corresponds to an entropy per particle of $0.63(2) k_B$.\\

\begin{figure}[t!]
	\centering
	\includegraphics[]{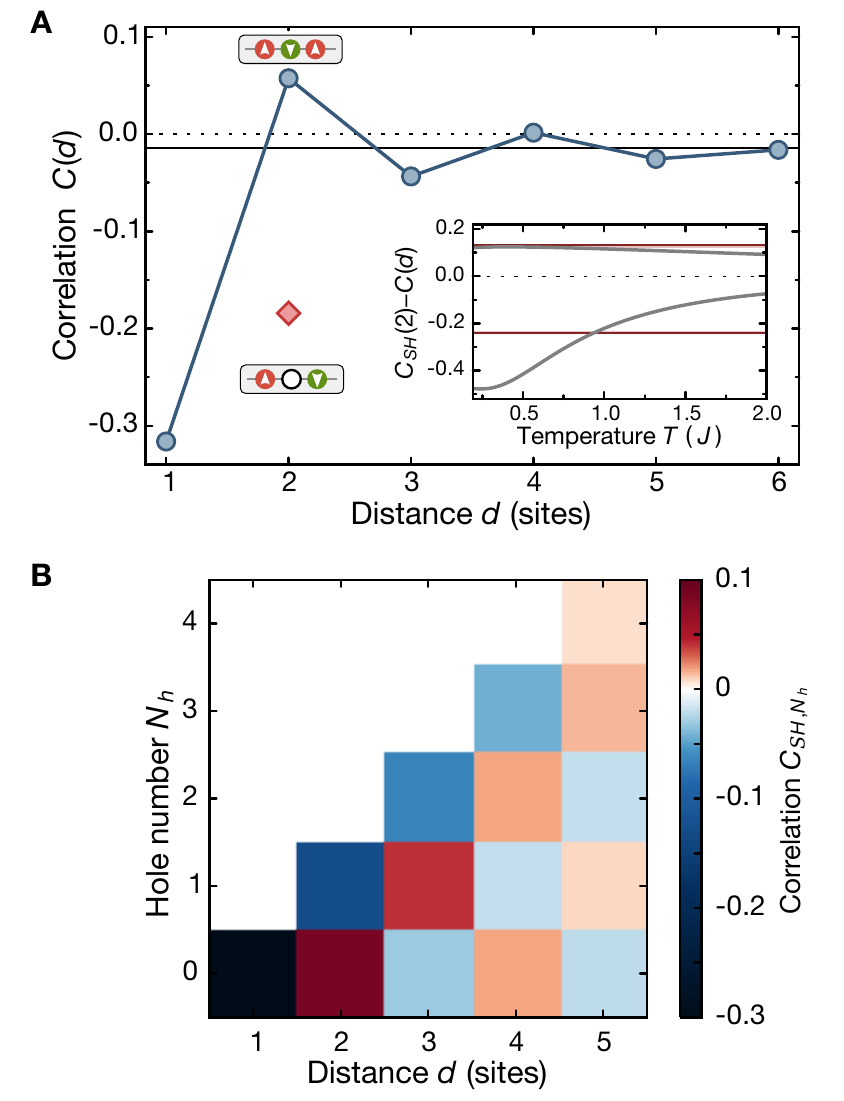}
\caption{
\textbf{Revealing the magnetic environment around holes.}
\textbf{(A)}  Connected two-point spin correlation function $C(d)$ analyzed on
occupied sites only (blue). The finite range AFM order without holes
asymptotically falls off with an exponential decay length of $1.3(2)$ sites. The spin correlations
at a distance of two sites switch sign in the presence of a hole as measured by
$C_{SH}(2)$ (red diamond) demonstrating an AFM environment surrounding the hole. The solid black
line indicates the finite-size offset~\cite{som}, the blue line is a guide to the eye
and statistical uncertainties are smaller than the symbol sizes. Inset:
Comparison of experimental values (red lines) of $C_{SH}(2)-C(1)$ (top) and
$C_{SH}(2)-C(2)$ (bottom) with finite temperature results from exact
diagonalization (gray curves). The light red shading is the systematic
correction due to a finite atom loss rate of up to 3\% during imaging.
\textbf{(B)} Amplitude of the correlation function $C_{SH,N_h}(d)$ as a
function of distance $d$ and the number of holes $N_h$ between the two spins
with the finite size offset subtracted. The parity of the AFM order flips with
every hole.
}
\end{figure}
To investigate the magnetic environment around a hole, we
calculate the conditional three-point spin-hole correlation function $C_{SH}(2)
= 4\,\langle\hat{S}^z_{i} \hat{S}^z_{i+2} \rangle_{\tiny\newmoon_i
\fullmoon_{i+1} \newmoon_{i+2}}$, where the symbols describe the condition that
the correlator is only evaluated on configurations with the sites $i$ and $i+2$
singly occupied and the middle site empty~\cite{som}. The correlator indeed
reveals anti-alignment of the spins around individual holes ($C_{SH}(2)<0$) and Fig.~2A
highlights the hole induced sign change by comparison to the standard two-point
correlator $$C(d)=4\left( \langle \hat{S}^z_i \hat{S}^z_{i+d}
\rangle_{\tiny\newmoon_i\newmoon_{i+d}} - \langle \hat{S}^z_i
\rangle_{\tiny\newmoon_i}\langle\hat{S}^z_{i+d} \rangle_{\tiny\newmoon_{i+d}}
\right)$$ To obtain unity filling, the latter was evaluated on a hole-free
subset of the data. The additional condition indicated by the symbols is
important in this paper as it removes the trivial $n^2$ density dependence of this
correlator, but has no effect here~\cite{som}.
The measured modulus of the correlation around a hole is
$|C_{SH}(2)|=0.184(4)$, considerably larger than $C(2)=0.057(3)$ and about half
of the next-neighbor value of $|C(1)|=0.316(2)$. At zero temperature for
$U/t\rightarrow\infty$ one expects $|C_{SH}(2)| = |C(1)|$, since the hole has no
effect on the magnetic alignment of its surrounding spins. For our interaction
strength, the measured difference agrees with exact diagonalization results at a
temperature $0.94(5)\,J$. These calculations take the
experimental fluctuations of the magnetization per chain into account. 
Due to finite size effects the correlation function shows a small offset at large
distances, for which we correct in the subsequent analysis throughout this
paper~\cite{som}.

The influence of larger doping on the spin order is revealed by studying
$C_{SH}(d)$ as a function of the number of holes between the two spins, that is
by evaluating $C_{SH,N_h}(d) = 4\,\langle\hat{S}^z_{i} \hat{S}^z_{i+d}
\rangle_{\tiny\newmoon_i \{\fullmoon\}_{N_h} \newmoon_{i+d}}$ with exactly
$N_h$ holes on the otherwise singly occupied string of length $d+1$. The
results of this analysis shown in Fig.~2B reveal a sign change of $C_{SH,N_h}$
at fixed distance $d$ for each newly introduced hole and antiferromagnetic
correlations versus distance for fixed hole number $N_h$. Thus, each hole indeed
corresponds to a flip of the antiferromagnetic parity. In a
thermodynamic ensemble, the hole number between the two measured spins
fluctuates, resulting in a weighted averaging over the alternating correlations
for different hole numbers. This directly explains the suppression of magnetic
correlations with hole doping (cf. Fig.~3A).\\

\begin{figure}[t!]
	\centering
	\includegraphics[]{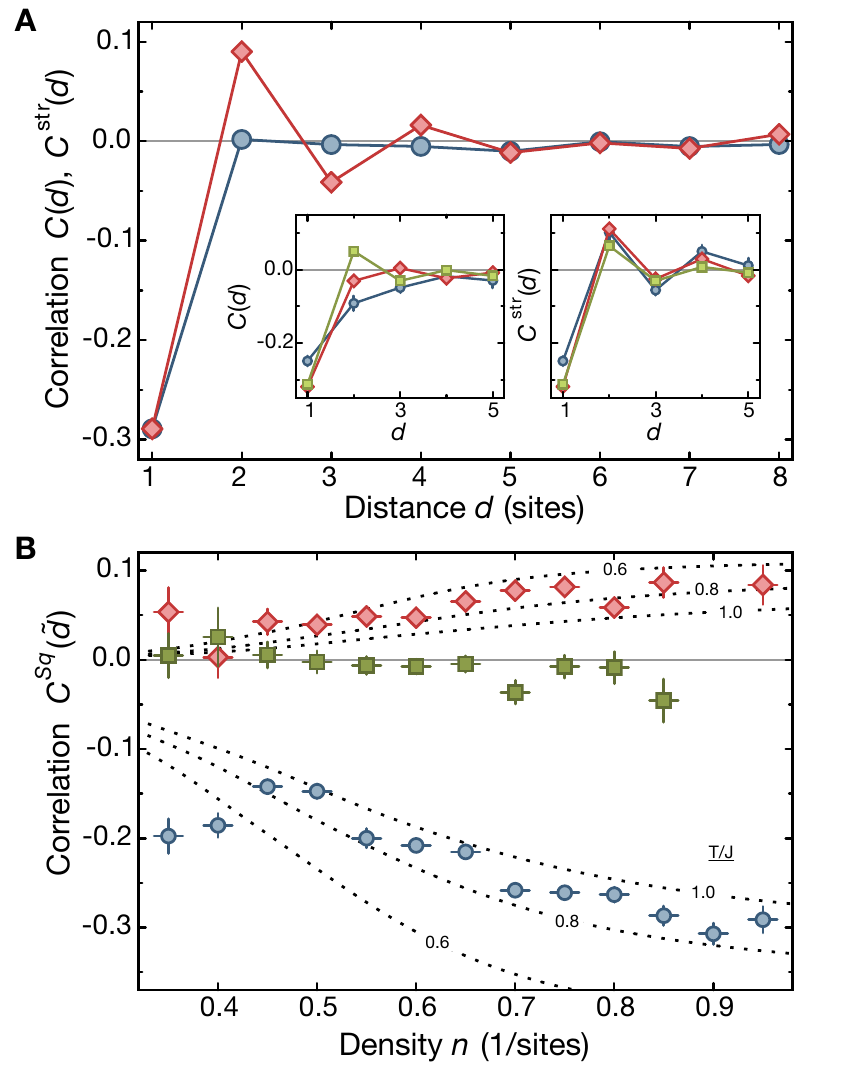}
\caption{
\textbf{Effect of hole doping on spin order.} \textbf{(A)} Comparison of the
spin correlation function (blue) $C(d)$  and the spin-string correlation
function (red) $C^{\textrm{str}}(d)$ averaged over all local densities in the
trap. The spin order is not visible with the conventional two-point spin correlator, but can be
revealed by disentangling spin and charge sector with the string correlator.
The extracted exponential decay length of $1.2(1)$ sites matches the one
extracted at unity filling (cf.~Fig.~1). The insets show the data binned by
density (bin widths $0.1$) for $\langle n\rangle=0.4$ (blue), $\langle
n\rangle=0.7$ (red) and $\langle n\rangle=1$ (green). Finite range AFM order in
the conventional correlator $C(d)$ is present at $\langle n\rangle=1$, while it
quickly gets suppressed when the system is doped away from half filling. At the
same time we observe an increasing periodicity of the two-point spin correlations with decreasing density
(left). In contrast, string correlations $C^{\textrm{str}}(d)$ only marginally
depend on density (right). Solid lines are guides to the eye. \textbf{(B)} Spin
correlation measured directly in squeezed space for $\tilde{d}=1$ (blue),
$\tilde{d}=2$ (red) and $\tilde{d}=3$ (green) as a function of density $n$ (bin widths $0.05$).
Dotted lines represent spin correlations $C(1)$ and $C(2)$ in the Heisenberg
model for temperatures $T/J=0.6,0.8,1.0$ obtained by exact diagonalization with
a coupling constant $J_{\textrm{eff}}(n)$. The correlation decreases with
increasing ratio $T/J_{\textrm{eff}}(n)$. All correlations shown are corrected
for the constant finite size offset~\cite{som}.
}
\end{figure}
The strong reduction of spin correlations due to hole fluctuations does not
imply the absence of magnetic order in the system, but rather suggests that it
is hidden by the fluctuations in the position of the atoms. This situation is
similar to the Haldane phase of spin-1 chains~\cite{haldane1983,haldane1983b, affleck1987,denNijs1989},
where fluctuating $\ket{0}$ spins hide correlations between the $\ket{\pm1}$
components leading to exponentially decaying local
correlators. The intrinsic AFM order is unveiled
by considering a non-local correlation function. By identifying double
occupancies and holes with spin $\ket{0}$ states, one can use the same
procedure to construct a string correlator that probes the underlying spin
order in the doped Hubbard chain~\cite{kruis2004}:
\begin{equation*}
C^{\textrm{str}}(d)=4 \left\langle \hat{S}^z_i \left(\prod_{j=1}^{d-1}(-1)^{(1-\hat{n}_{i+j})}\right)
\hat{S}^z_{i+d}\right\rangle_{\tiny\newmoon_i\newmoon_{i+d}}
\end{equation*}
This string correlator takes the antiferromagnetic parity flips into account by
a corresponding sign flip for each hole (cf.~Fig.~1C). The unique ability to
detect the spin and density locally on single images~\cite{boll2016} enables
the direct measurement of the string correlator $C^{\textrm{\textrm{str}}}(d)$
for different densities. The dependence of the string correlator on distance
reported on Fig.~3A is in stark constrast with the standard two-point spin
correlation function $C(d)$. While $C(d)$ quickly vanishes when the system is
doped away from half filling, staggered correlations at distances up to four
sites are detected with the string correlator $C^{\textrm{str}}(d)$ (Fig.~3A). 
When analyzing the data in regions of fixed density, we additionally observe 
 an increasing periodicity of the AFM correlations with decreasing
density~\cite{giamarchi2004}. The amplitude of the string correlations, on the
other hand, even slightly increase in magnitude at a given real space distance
$d$ which we attribute to the decreasing distance $\tilde{d}\sim n d$ in squeezed
space~\cite{som}.

An analysis of the correlations directly in squeezed space is also possible
with the quantum gas microscope by removing the empty and doubly occupied sites
in the analysis before evaluating the standard two-point correlator $C(d)$.
This corresponds to a weighted summation along the diagonals of Fig.~2B, and thus
mixes events that had different distances in real space. Similar to the string
correlator, the squeezed space analysis~(Fig.~3B) reveals the finite-range hidden
antiferromagnetic order. A quantitative comparison to a
Heisenberg model with renormalized coupling $J_{\textrm{eff}}(n)$, that decreases with doping, agrees well at a temperature of $T=0.87(2)~J$, which demonstrates that the concept
of squeezed space can be successfully applied even away from the
$U/t\rightarrow\infty$ limit~\cite{imambekov2012}. Here, $J_{\textrm{eff}}$ was determined
independently from the microscopic parameters of the Hubbard model~\cite{som}.
The discrepancy between theory and experiments at densities below $0.45$ might arise
from adiabatic cooling when decreasing the density during the preparation of the chains.

\begin{figure*}[t!]
	\centering
	\includegraphics[]{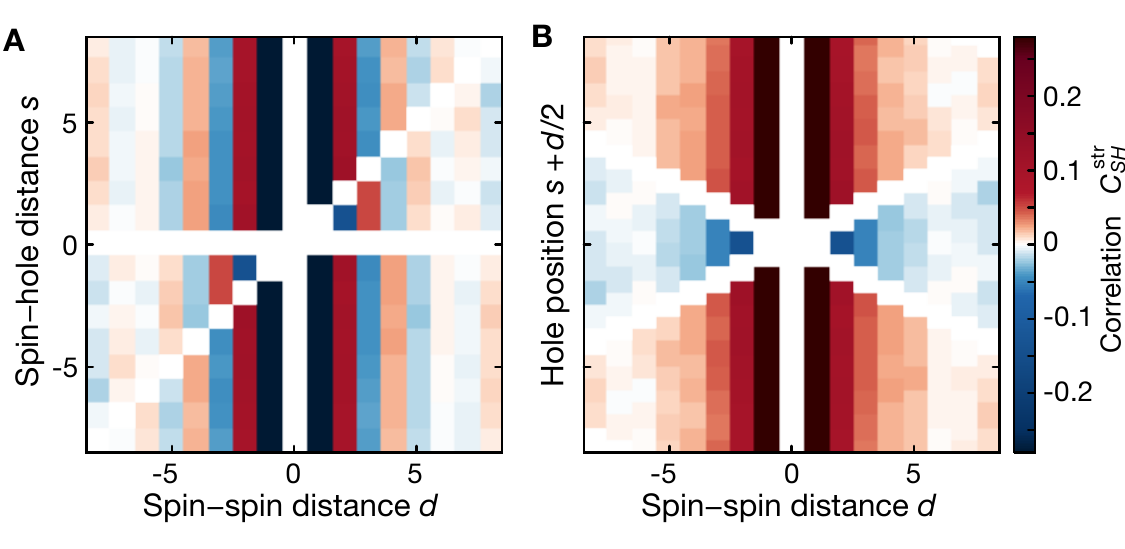}

\caption{
\textbf{Single holes as domain walls for the AFM order.} \textbf{(A)} Tailored
string correlator $C^{\textrm{str}}_{SH}(d,s)$ measuring the effect of a
single hole on the doped Hubbard-chain. As expected for separated spin and
charge sectors, the correlations are independent of the distance $s$ between the hole and
the spin, except for the opposite sign when the hole sits in between the
two spins at relative distance $d$. In addition, there is a dynamic picture to
the measurements shown here. Interpreting the vertical axis as time, one
obtains the picture of a delocalized hole freely propagating through an
antiferromagnetic background. The correlator  $C^{\textrm{str}}_{SH}(d,s)$ is
set to zero whenever two operators are evaluated at the same site. \textbf{(B)}
Rectified correlator $(-1)^d\,C^{\rm str}_{SH}(d,s)$ with hole position
referenced to the string center. The hole associated AFM parity flips are
directly visible by the different domains. The expected parity is
observed consistently for spin-spin distances of up to eight sites. 
}
\end{figure*}

In order to further confirm the independence of the spin and density sectors,
we define a tailored string correlator
 $$C_{SH}^{\textrm{str}}(d,s)=4\,\left\langle
\hat{S}^z_i  \left(\prod_{\substack{j=1,j\neq s}}^{d-1}
(-1)^{(1-\hat{n}_{i+j})}\right) \hat{S}^z_{i+d} \right\rangle_{ \tiny\newmoon_i
\fullmoon_{i+s} \newmoon_{i+d}}$$ which isolates the effect of a single hole at distance $s$
from the first spin
independent of the density. Here, the effect of extra charge fluctuations is
taken care of by inserting string correlators around the hole. For a system with a single hole this correlator is
identical to the three-point correlation function introduced before
$C_{SH}^{\textrm{str}}$ = $C_{SH}$. The dependence of $C_{SH}^{\textrm{str}}$
on the spin separation $d$ and the position of the hole in the string $s$ is
shown in Fig.~4A. For $s = 0$ and $s = d$ the hole crosses one of the two
spins, which causes the previously discussed AFM parity flip, while the
correlation signal is almost independent of the position of the hole between
the two spins. This observation emphasizes spin-charge separation by the
absence of polaron-like effects, which would result in a local change of the
spin correlations around the hole. The rectified
correlator $(-1)^d C_{SH}^{\textrm{str}}(d,s)$ in Fig.~4B highlights the two
domains of opposite AFM parity, demonstrating that the hole acts as a domain
wall for the magnetic order \cite{zaanen1998}. To emphasize the symmetries of the three-point
correlator, the position of the hole is measured here relative to the center of
mass of the two spins.

Through the analysis of various local and non-local correlation functions our
measurements revealed striking equilibrium signatures of spin-charge separation
in one-dimensional Hubbard chains. An interesting extension of this work would be the
detection of dynamic signatures of spin-charge separation in quench experiments
through the measurements of different spin and charge velocities~\cite{recati2003,kollath2005}. In higher dimensions the experimental evaluation of non-local correlations in synthetic
hole doped antiferromagnetic materials is also of prime interest for the investigation
of exotic many-body phases relevant to high temperature
superconductors~\cite{carlson2004,zhang2002}. The extension to two-dimensional frustrated quantum magnets would,
for example, enable the detection of deconfined criticality through Wilson loops~\cite{senthil2004}. Hence, our experiments mark a first step towards experimental studies of emergent gauge structures and topological order~\cite{wen2004}.

\begin{flushleft}
  \textbf{Acknowledgements} We acknowledge J. Koepsell and J. Vijayan for a critical reading of the manuscript, A. Sterdyniak and M. Zvonarev for useful discussions, and financial support by MPG and EU (UQUAM). FG and ED acknowledge support from Harvard-MIT CUA, NSF Grant No. DMR-1308435, the Moore Foundation,  AFOSR Quantum Simulation MURI, AFOSR MURI Photonic Quantum Matter.
\end{flushleft}


\bibliography{stringcorr,blochgrouppapers}

\begin{thebibliography}{42}%
\makeatletter
\providecommand \@ifxundefined [1]{%
 \@ifx{#1\undefined}
}%
\providecommand \@ifnum [1]{%
 \ifnum #1\expandafter \@firstoftwo
 \else \expandafter \@secondoftwo
 \fi
}%
\providecommand \@ifx [1]{%
 \ifx #1\expandafter \@firstoftwo
 \else \expandafter \@secondoftwo
 \fi
}%
\providecommand \natexlab [1]{#1}%
\providecommand \enquote  [1]{``#1''}%
\providecommand \bibnamefont  [1]{#1}%
\providecommand \bibfnamefont [1]{#1}%
\providecommand \citenamefont [1]{#1}%
\providecommand \href@noop [0]{\@secondoftwo}%
\providecommand \href [0]{\begingroup \@sanitize@url \@href}%
\providecommand \@href[1]{\@@startlink{#1}\@@href}%
\providecommand \@@href[1]{\endgroup#1\@@endlink}%
\providecommand \@sanitize@url [0]{\catcode `\\12\catcode `\$12\catcode
  `\&12\catcode `\#12\catcode `\^12\catcode `\_12\catcode `\%12\relax}%
\providecommand \@@startlink[1]{}%
\providecommand \@@endlink[0]{}%
\providecommand \url  [0]{\begingroup\@sanitize@url \@url }%
\providecommand \@url [1]{\endgroup\@href {#1}{\urlprefix }}%
\providecommand \urlprefix  [0]{URL }%
\providecommand \Eprint [0]{\href }%
\providecommand \doibase [0]{http://dx.doi.org/}%
\providecommand \selectlanguage [0]{\@gobble}%
\providecommand \bibinfo  [0]{\@secondoftwo}%
\providecommand \bibfield  [0]{\@secondoftwo}%
\providecommand \translation [1]{[#1]}%
\providecommand \BibitemOpen [0]{}%
\providecommand \bibitemStop [0]{}%
\providecommand \bibitemNoStop [0]{.\EOS\space}%
\providecommand \EOS [0]{\spacefactor3000\relax}%
\providecommand \BibitemShut  [1]{\csname bibitem#1\endcsname}%
\let\auto@bib@innerbib\@empty
\bibitem [{\citenamefont {Auerbach}(1994)}]{auerbach1994}%
  \BibitemOpen
  \bibfield  {author} {\bibinfo {author} {\bibfnamefont {A.}~\bibnamefont
  {Auerbach}},\ }\href@noop {} {\emph {\bibinfo {title} {Interacting
  {{Electrons}} and {{Quantum Magnetism}}}}}\ (\bibinfo  {publisher} {{Springer
  Science \& Business Media}},\ \bibinfo {year} {1994})\BibitemShut {NoStop}%
\bibitem [{\citenamefont {Lee}\ \emph {et~al.}(2006)\citenamefont {Lee},
  \citenamefont {Nagaosa},\ and\ \citenamefont {Wen}}]{Lee2006}%
  \BibitemOpen
  \bibfield  {author} {\bibinfo {author} {\bibfnamefont {P.~A.}\ \bibnamefont
  {Lee}}, \bibinfo {author} {\bibfnamefont {N.}~\bibnamefont {Nagaosa}}, \ and\
  \bibinfo {author} {\bibfnamefont {X.-G.}\ \bibnamefont {Wen}},\ }\href
  {\doibase 10.1103/RevModPhys.78.17} {\bibfield  {journal} {\bibinfo
  {journal} {Rev. Mod. Phys.}\ }\textbf {\bibinfo {volume} {78}},\ \bibinfo
  {pages} {17} (\bibinfo {year} {2006})}\BibitemShut {NoStop}%
\bibitem [{\citenamefont {Giamarchi}(2004)}]{giamarchi2004}%
  \BibitemOpen
  \bibfield  {author} {\bibinfo {author} {\bibfnamefont {T.}~\bibnamefont
  {Giamarchi}},\ }\href@noop {} {\emph {\bibinfo {title} {Quantum {{Physics}}
  in {{One Dimension}}}}}\ (\bibinfo  {publisher} {{Clarendon Press}},\
  \bibinfo {year} {2004})\BibitemShut {NoStop}%
\bibitem [{\citenamefont {Kim}\ \emph {et~al.}(1996)\citenamefont {Kim},
  \citenamefont {Matsuura}, \citenamefont {Shen}, \citenamefont {Motoyama},
  \citenamefont {Eisaki}, \citenamefont {Uchida}, \citenamefont {Tohyama},\
  and\ \citenamefont {Maekawa}}]{kim1996}%
  \BibitemOpen
  \bibfield  {author} {\bibinfo {author} {\bibfnamefont {C.}~\bibnamefont
  {Kim}}, \bibinfo {author} {\bibfnamefont {A.~Y.}\ \bibnamefont {Matsuura}},
  \bibinfo {author} {\bibfnamefont {Z.-X.}\ \bibnamefont {Shen}}, \bibinfo
  {author} {\bibfnamefont {N.}~\bibnamefont {Motoyama}}, \bibinfo {author}
  {\bibfnamefont {H.}~\bibnamefont {Eisaki}}, \bibinfo {author} {\bibfnamefont
  {S.}~\bibnamefont {Uchida}}, \bibinfo {author} {\bibfnamefont
  {T.}~\bibnamefont {Tohyama}}, \ and\ \bibinfo {author} {\bibfnamefont
  {S.}~\bibnamefont {Maekawa}},\ }\href {\doibase 10.1103/PhysRevLett.77.4054}
  {\bibfield  {journal} {\bibinfo  {journal} {Phys. Rev. Lett.}\ }\textbf
  {\bibinfo {volume} {77}},\ \bibinfo {pages} {4054} (\bibinfo {year}
  {1996})}\BibitemShut {NoStop}%
\bibitem [{\citenamefont {Segovia}\ \emph {et~al.}(1999)\citenamefont
  {Segovia}, \citenamefont {Purdie}, \citenamefont {Hengsberger},\ and\
  \citenamefont {Baer}}]{segovia1999}%
  \BibitemOpen
  \bibfield  {author} {\bibinfo {author} {\bibfnamefont {P.}~\bibnamefont
  {Segovia}}, \bibinfo {author} {\bibfnamefont {D.}~\bibnamefont {Purdie}},
  \bibinfo {author} {\bibfnamefont {M.}~\bibnamefont {Hengsberger}}, \ and\
  \bibinfo {author} {\bibfnamefont {Y.}~\bibnamefont {Baer}},\ }\href {\doibase
  10.1038/990052} {\bibfield  {journal} {\bibinfo  {journal} {Nature}\ }\textbf
  {\bibinfo {volume} {402}},\ \bibinfo {pages} {504} (\bibinfo {year}
  {1999})}\BibitemShut {NoStop}%
\bibitem [{\citenamefont {Kim}\ \emph {et~al.}(2006)\citenamefont {Kim},
  \citenamefont {Koh}, \citenamefont {Rotenberg}, \citenamefont {Oh},
  \citenamefont {Eisaki}, \citenamefont {Motoyama}, \citenamefont {Uchida},
  \citenamefont {Tohyama}, \citenamefont {Maekawa}, \citenamefont {Shen},\ and\
  \citenamefont {Kim}}]{kim2006}%
  \BibitemOpen
  \bibfield  {author} {\bibinfo {author} {\bibfnamefont {B.~J.}\ \bibnamefont
  {Kim}}, \bibinfo {author} {\bibfnamefont {H.}~\bibnamefont {Koh}}, \bibinfo
  {author} {\bibfnamefont {E.}~\bibnamefont {Rotenberg}}, \bibinfo {author}
  {\bibfnamefont {S.-J.}\ \bibnamefont {Oh}}, \bibinfo {author} {\bibfnamefont
  {H.}~\bibnamefont {Eisaki}}, \bibinfo {author} {\bibfnamefont
  {N.}~\bibnamefont {Motoyama}}, \bibinfo {author} {\bibfnamefont
  {S.}~\bibnamefont {Uchida}}, \bibinfo {author} {\bibfnamefont
  {T.}~\bibnamefont {Tohyama}}, \bibinfo {author} {\bibfnamefont
  {S.}~\bibnamefont {Maekawa}}, \bibinfo {author} {\bibfnamefont {Z.-X.}\
  \bibnamefont {Shen}}, \ and\ \bibinfo {author} {\bibfnamefont
  {C.}~\bibnamefont {Kim}},\ }\href {\doibase 10.1038/nphys316} {\bibfield
  {journal} {\bibinfo  {journal} {Nat. Phys.}\ }\textbf {\bibinfo {volume}
  {2}},\ \bibinfo {pages} {397} (\bibinfo {year} {2006})}\BibitemShut {NoStop}%
\bibitem [{\citenamefont {Auslaender}\ \emph {et~al.}(2005)\citenamefont
  {Auslaender}, \citenamefont {Steinberg}, \citenamefont {Yacoby},
  \citenamefont {Tserkovnyak}, \citenamefont {Halperin}, \citenamefont
  {Baldwin}, \citenamefont {Pfeiffer},\ and\ \citenamefont
  {West}}]{auslaender2005}%
  \BibitemOpen
  \bibfield  {author} {\bibinfo {author} {\bibfnamefont {O.~M.}\ \bibnamefont
  {Auslaender}}, \bibinfo {author} {\bibfnamefont {H.}~\bibnamefont
  {Steinberg}}, \bibinfo {author} {\bibfnamefont {A.}~\bibnamefont {Yacoby}},
  \bibinfo {author} {\bibfnamefont {Y.}~\bibnamefont {Tserkovnyak}}, \bibinfo
  {author} {\bibfnamefont {B.~I.}\ \bibnamefont {Halperin}}, \bibinfo {author}
  {\bibfnamefont {K.~W.}\ \bibnamefont {Baldwin}}, \bibinfo {author}
  {\bibfnamefont {L.~N.}\ \bibnamefont {Pfeiffer}}, \ and\ \bibinfo {author}
  {\bibfnamefont {K.~W.}\ \bibnamefont {West}},\ }\href {\doibase
  10.1126/science.1107821} {\bibfield  {journal} {\bibinfo  {journal}
  {Science}\ }\textbf {\bibinfo {volume} {308}},\ \bibinfo {pages} {88}
  (\bibinfo {year} {2005})}\BibitemShut {NoStop}%
\bibitem [{\citenamefont {Jompol}\ \emph {et~al.}(2009)\citenamefont {Jompol},
  \citenamefont {Ford}, \citenamefont {Griffiths}, \citenamefont {Farrer},
  \citenamefont {Jones}, \citenamefont {Anderson}, \citenamefont {Ritchie},
  \citenamefont {Silk},\ and\ \citenamefont {Schofield}}]{jompol2009}%
  \BibitemOpen
  \bibfield  {author} {\bibinfo {author} {\bibfnamefont {Y.}~\bibnamefont
  {Jompol}}, \bibinfo {author} {\bibfnamefont {C.~J.~B.}\ \bibnamefont {Ford}},
  \bibinfo {author} {\bibfnamefont {J.~P.}\ \bibnamefont {Griffiths}}, \bibinfo
  {author} {\bibfnamefont {I.}~\bibnamefont {Farrer}}, \bibinfo {author}
  {\bibfnamefont {G.~a.~C.}\ \bibnamefont {Jones}}, \bibinfo {author}
  {\bibfnamefont {D.}~\bibnamefont {Anderson}}, \bibinfo {author}
  {\bibfnamefont {D.~A.}\ \bibnamefont {Ritchie}}, \bibinfo {author}
  {\bibfnamefont {T.~W.}\ \bibnamefont {Silk}}, \ and\ \bibinfo {author}
  {\bibfnamefont {A.~J.}\ \bibnamefont {Schofield}},\ }\href {\doibase
  10.1126/science.1171769} {\bibfield  {journal} {\bibinfo  {journal}
  {Science}\ }\textbf {\bibinfo {volume} {325}},\ \bibinfo {pages} {597}
  (\bibinfo {year} {2009})}\BibitemShut {NoStop}%
\bibitem [{\citenamefont {{den Nijs}}\ and\ \citenamefont
  {Rommelse}(1989)}]{denNijs1989}%
  \BibitemOpen
  \bibfield  {author} {\bibinfo {author} {\bibfnamefont {M.}~\bibnamefont {{den
  Nijs}}}\ and\ \bibinfo {author} {\bibfnamefont {K.}~\bibnamefont
  {Rommelse}},\ }\href {\doibase 10.1103/PhysRevB.40.4709} {\bibfield
  {journal} {\bibinfo  {journal} {Phys. Rev. B}\ }\textbf {\bibinfo {volume}
  {40}},\ \bibinfo {pages} {4709} (\bibinfo {year} {1989})}\BibitemShut
  {NoStop}%
\bibitem [{\citenamefont {Kennedy}\ and\ \citenamefont
  {Tasaki}(1992)}]{kennedy1992b}%
  \BibitemOpen
  \bibfield  {author} {\bibinfo {author} {\bibfnamefont {T.}~\bibnamefont
  {Kennedy}}\ and\ \bibinfo {author} {\bibfnamefont {H.}~\bibnamefont
  {Tasaki}},\ }\href {\doibase 10.1103/PhysRevB.45.304} {\bibfield  {journal}
  {\bibinfo  {journal} {Phys. Rev. B}\ }\textbf {\bibinfo {volume} {45}},\
  \bibinfo {pages} {304} (\bibinfo {year} {1992})}\BibitemShut {NoStop}%
\bibitem [{\citenamefont {Kruis}\ \emph
  {et~al.}(2004{\natexlab{a}})\citenamefont {Kruis}, \citenamefont {McCulloch},
  \citenamefont {Nussinov},\ and\ \citenamefont {Zaanen}}]{kruis2004a}%
  \BibitemOpen
  \bibfield  {author} {\bibinfo {author} {\bibfnamefont {H.~V.}\ \bibnamefont
  {Kruis}}, \bibinfo {author} {\bibfnamefont {I.~P.}\ \bibnamefont
  {McCulloch}}, \bibinfo {author} {\bibfnamefont {Z.}~\bibnamefont {Nussinov}},
  \ and\ \bibinfo {author} {\bibfnamefont {J.}~\bibnamefont {Zaanen}},\ }\href
  {\doibase 10.1209/epl/i2003-10114-3} {\bibfield  {journal} {\bibinfo
  {journal} {Europhys. Lett. (EPL)}\ }\textbf {\bibinfo {volume} {65}},\
  \bibinfo {pages} {512} (\bibinfo {year} {2004}{\natexlab{a}})}\BibitemShut
  {NoStop}%
\bibitem [{\citenamefont {Haldane}(1983{\natexlab{a}})}]{haldane1983}%
  \BibitemOpen
  \bibfield  {author} {\bibinfo {author} {\bibfnamefont {F.}~\bibnamefont
  {Haldane}},\ }\href {\doibase 10.1016/0375-9601(83)90631-x} {\bibfield
  {journal} {\bibinfo  {journal} {Physics Letters A}\ }\textbf {\bibinfo
  {volume} {93}},\ \bibinfo {pages} {464} (\bibinfo {year}
  {1983}{\natexlab{a}})}\BibitemShut {NoStop}%
\bibitem [{\citenamefont {Haldane}(1983{\natexlab{b}})}]{haldane1983b}%
  \BibitemOpen
  \bibfield  {author} {\bibinfo {author} {\bibfnamefont {F.~D.~M.}\
  \bibnamefont {Haldane}},\ }\href {\doibase 10.1103/PhysRevLett.50.1153}
  {\bibfield  {journal} {\bibinfo  {journal} {Phys. Rev. Lett.}\ }\textbf
  {\bibinfo {volume} {50}},\ \bibinfo {pages} {1153} (\bibinfo {year}
  {1983}{\natexlab{b}})}\BibitemShut {NoStop}%
\bibitem [{\citenamefont {Affleck}\ \emph {et~al.}(1987)\citenamefont
  {Affleck}, \citenamefont {Kennedy}, \citenamefont {Lieb},\ and\ \citenamefont
  {Tasaki}}]{affleck1987}%
  \BibitemOpen
  \bibfield  {author} {\bibinfo {author} {\bibfnamefont {I.}~\bibnamefont
  {Affleck}}, \bibinfo {author} {\bibfnamefont {T.}~\bibnamefont {Kennedy}},
  \bibinfo {author} {\bibfnamefont {E.~H.}\ \bibnamefont {Lieb}}, \ and\
  \bibinfo {author} {\bibfnamefont {H.}~\bibnamefont {Tasaki}},\ }\href
  {\doibase 10.1103/PhysRevLett.59.799} {\bibfield  {journal} {\bibinfo
  {journal} {Phys. Rev. Lett.}\ }\textbf {\bibinfo {volume} {59}},\ \bibinfo
  {pages} {799} (\bibinfo {year} {1987})}\BibitemShut {NoStop}%
\bibitem [{\citenamefont {Ogata}\ and\ \citenamefont
  {Shiba}(1990)}]{ogata1990}%
  \BibitemOpen
  \bibfield  {author} {\bibinfo {author} {\bibfnamefont {M.}~\bibnamefont
  {Ogata}}\ and\ \bibinfo {author} {\bibfnamefont {H.}~\bibnamefont {Shiba}},\
  }\href {\doibase 10.1103/PhysRevB.41.2326} {\bibfield  {journal} {\bibinfo
  {journal} {Phys. Rev. B}\ }\textbf {\bibinfo {volume} {41}},\ \bibinfo
  {pages} {2326} (\bibinfo {year} {1990})}\BibitemShut {NoStop}%
\bibitem [{\citenamefont {Ren}\ and\ \citenamefont {Anderson}(1993)}]{ren1993}%
  \BibitemOpen
  \bibfield  {author} {\bibinfo {author} {\bibfnamefont {Y.}~\bibnamefont
  {Ren}}\ and\ \bibinfo {author} {\bibfnamefont {P.~W.}\ \bibnamefont
  {Anderson}},\ }\href {\doibase 10.1103/PhysRevB.48.16662} {\bibfield
  {journal} {\bibinfo  {journal} {Phys. Rev. B}\ }\textbf {\bibinfo {volume}
  {48}},\ \bibinfo {pages} {16662} (\bibinfo {year} {1993})}\BibitemShut
  {NoStop}%
\bibitem [{\citenamefont {Zaanen}\ \emph {et~al.}(2001)\citenamefont {Zaanen},
  \citenamefont {Osman}, \citenamefont {Kruis}, \citenamefont {Nussinov},\ and\
  \citenamefont {Tworzydlo}}]{zaanen2001}%
  \BibitemOpen
  \bibfield  {author} {\bibinfo {author} {\bibfnamefont {J.}~\bibnamefont
  {Zaanen}}, \bibinfo {author} {\bibfnamefont {O.~Y.}\ \bibnamefont {Osman}},
  \bibinfo {author} {\bibfnamefont {H.~V.}\ \bibnamefont {Kruis}}, \bibinfo
  {author} {\bibfnamefont {Z.}~\bibnamefont {Nussinov}}, \ and\ \bibinfo
  {author} {\bibfnamefont {J.}~\bibnamefont {Tworzydlo}},\ }\href {\doibase
  10.1080/13642810108208566} {\bibfield  {journal} {\bibinfo  {journal}
  {Philos. Mag. B}\ }\textbf {\bibinfo {volume} {81}},\ \bibinfo {pages} {1485}
  (\bibinfo {year} {2001})}\BibitemShut {NoStop}%
\bibitem [{\citenamefont {Kruis}\ \emph
  {et~al.}(2004{\natexlab{b}})\citenamefont {Kruis}, \citenamefont {McCulloch},
  \citenamefont {Nussinov},\ and\ \citenamefont {Zaanen}}]{kruis2004}%
  \BibitemOpen
  \bibfield  {author} {\bibinfo {author} {\bibfnamefont {H.~V.}\ \bibnamefont
  {Kruis}}, \bibinfo {author} {\bibfnamefont {I.~P.}\ \bibnamefont
  {McCulloch}}, \bibinfo {author} {\bibfnamefont {Z.}~\bibnamefont {Nussinov}},
  \ and\ \bibinfo {author} {\bibfnamefont {J.}~\bibnamefont {Zaanen}},\ }\href
  {\doibase 10.1103/PhysRevB.70.075109} {\bibfield  {journal} {\bibinfo
  {journal} {Phys. Rev. B}\ }\textbf {\bibinfo {volume} {70}},\ \bibinfo
  {pages} {075109} (\bibinfo {year} {2004}{\natexlab{b}})}\BibitemShut
  {NoStop}%
\bibitem [{\citenamefont {Haller}\ \emph {et~al.}(2015)\citenamefont {Haller},
  \citenamefont {Hudson}, \citenamefont {Kelly}, \citenamefont {Cotta},
  \citenamefont {Peaudecerf}, \citenamefont {Bruce},\ and\ \citenamefont
  {Kuhr}}]{haller2015}%
  \BibitemOpen
  \bibfield  {author} {\bibinfo {author} {\bibfnamefont {E.}~\bibnamefont
  {Haller}}, \bibinfo {author} {\bibfnamefont {J.}~\bibnamefont {Hudson}},
  \bibinfo {author} {\bibfnamefont {A.}~\bibnamefont {Kelly}}, \bibinfo
  {author} {\bibfnamefont {D.~A.}\ \bibnamefont {Cotta}}, \bibinfo {author}
  {\bibfnamefont {B.}~\bibnamefont {Peaudecerf}}, \bibinfo {author}
  {\bibfnamefont {G.~D.}\ \bibnamefont {Bruce}}, \ and\ \bibinfo {author}
  {\bibfnamefont {S.}~\bibnamefont {Kuhr}},\ }\href {\doibase
  10.1038/nphys3403} {\bibfield  {journal} {\bibinfo  {journal} {Nat. Phys.}\
  }\textbf {\bibinfo {volume} {11}},\ \bibinfo {pages} {738} (\bibinfo {year}
  {2015})}\BibitemShut {NoStop}%
\bibitem [{\citenamefont {Cheuk}\ \emph {et~al.}(2016)\citenamefont {Cheuk},
  \citenamefont {Nichols}, \citenamefont {Lawrence}, \citenamefont {Okan},
  \citenamefont {Zhang}, \citenamefont {Khatami}, \citenamefont {Trivedi},
  \citenamefont {Paiva}, \citenamefont {Rigol},\ and\ \citenamefont
  {Zwierlein}}]{cheuk2016}%
  \BibitemOpen
  \bibfield  {author} {\bibinfo {author} {\bibfnamefont {L.~W.}\ \bibnamefont
  {Cheuk}}, \bibinfo {author} {\bibfnamefont {M.~A.}\ \bibnamefont {Nichols}},
  \bibinfo {author} {\bibfnamefont {K.~R.}\ \bibnamefont {Lawrence}}, \bibinfo
  {author} {\bibfnamefont {M.}~\bibnamefont {Okan}}, \bibinfo {author}
  {\bibfnamefont {H.}~\bibnamefont {Zhang}}, \bibinfo {author} {\bibfnamefont
  {E.}~\bibnamefont {Khatami}}, \bibinfo {author} {\bibfnamefont
  {N.}~\bibnamefont {Trivedi}}, \bibinfo {author} {\bibfnamefont
  {T.}~\bibnamefont {Paiva}}, \bibinfo {author} {\bibfnamefont
  {M.}~\bibnamefont {Rigol}}, \ and\ \bibinfo {author} {\bibfnamefont {M.~W.}\
  \bibnamefont {Zwierlein}},\ }\href {\doibase 10.1126/science.aag3349}
  {\bibfield  {journal} {\bibinfo  {journal} {Science}\ }\textbf {\bibinfo
  {volume} {353}},\ \bibinfo {pages} {1260} (\bibinfo {year}
  {2016})}\BibitemShut {NoStop}%
\bibitem [{\citenamefont {Parsons}\ \emph {et~al.}(2016)\citenamefont
  {Parsons}, \citenamefont {Mazurenko}, \citenamefont {Chiu}, \citenamefont
  {Ji}, \citenamefont {Greif},\ and\ \citenamefont {Greiner}}]{parsons2016}%
  \BibitemOpen
  \bibfield  {author} {\bibinfo {author} {\bibfnamefont {M.~F.}\ \bibnamefont
  {Parsons}}, \bibinfo {author} {\bibfnamefont {A.}~\bibnamefont {Mazurenko}},
  \bibinfo {author} {\bibfnamefont {C.~S.}\ \bibnamefont {Chiu}}, \bibinfo
  {author} {\bibfnamefont {G.}~\bibnamefont {Ji}}, \bibinfo {author}
  {\bibfnamefont {D.}~\bibnamefont {Greif}}, \ and\ \bibinfo {author}
  {\bibfnamefont {M.}~\bibnamefont {Greiner}},\ }\href {\doibase
  10.1126/science.aag1430} {\bibfield  {journal} {\bibinfo  {journal}
  {Science}\ }\textbf {\bibinfo {volume} {353}},\ \bibinfo {pages} {1253}
  (\bibinfo {year} {2016})}\BibitemShut {NoStop}%
\bibitem [{\citenamefont {Omran}\ \emph {et~al.}(2015)\citenamefont {Omran},
  \citenamefont {Boll}, \citenamefont {Hilker}, \citenamefont {Kleinlein},
  \citenamefont {Salomon}, \citenamefont {Bloch},\ and\ \citenamefont
  {Gross}}]{omran2015}%
  \BibitemOpen
  \bibfield  {author} {\bibinfo {author} {\bibfnamefont {A.}~\bibnamefont
  {Omran}}, \bibinfo {author} {\bibfnamefont {M.}~\bibnamefont {Boll}},
  \bibinfo {author} {\bibfnamefont {T.~A.}\ \bibnamefont {Hilker}}, \bibinfo
  {author} {\bibfnamefont {K.}~\bibnamefont {Kleinlein}}, \bibinfo {author}
  {\bibfnamefont {G.}~\bibnamefont {Salomon}}, \bibinfo {author} {\bibfnamefont
  {I.}~\bibnamefont {Bloch}}, \ and\ \bibinfo {author} {\bibfnamefont
  {C.}~\bibnamefont {Gross}},\ }\href {\doibase 10.1103/PhysRevLett.115.263001}
  {\bibfield  {journal} {\bibinfo  {journal} {Phys. Rev. Lett.}\ }\textbf
  {\bibinfo {volume} {115}},\ \bibinfo {pages} {263001} (\bibinfo {year}
  {2015})}\BibitemShut {NoStop}%
\bibitem [{\citenamefont {Edge}\ \emph {et~al.}(2015)\citenamefont {Edge},
  \citenamefont {Anderson}, \citenamefont {Jervis}, \citenamefont {McKay},
  \citenamefont {Day}, \citenamefont {Trotzky},\ and\ \citenamefont
  {Thywissen}}]{edge2015}%
  \BibitemOpen
  \bibfield  {author} {\bibinfo {author} {\bibfnamefont {G.~J.~A.}\
  \bibnamefont {Edge}}, \bibinfo {author} {\bibfnamefont {R.}~\bibnamefont
  {Anderson}}, \bibinfo {author} {\bibfnamefont {D.}~\bibnamefont {Jervis}},
  \bibinfo {author} {\bibfnamefont {D.~C.}\ \bibnamefont {McKay}}, \bibinfo
  {author} {\bibfnamefont {R.}~\bibnamefont {Day}}, \bibinfo {author}
  {\bibfnamefont {S.}~\bibnamefont {Trotzky}}, \ and\ \bibinfo {author}
  {\bibfnamefont {J.~H.}\ \bibnamefont {Thywissen}},\ }\href {\doibase
  10.1103/PhysRevA.92.063406} {\bibfield  {journal} {\bibinfo  {journal} {Phys.
  Rev. A}\ }\textbf {\bibinfo {volume} {92}},\ \bibinfo {pages} {063406}
  (\bibinfo {year} {2015})}\BibitemShut {NoStop}%
\bibitem [{\citenamefont {{Brown}}\ \emph {et~al.}(2016)\citenamefont
  {{Brown}}, \citenamefont {{Mitra}}, \citenamefont {{Guardado-Sanchez}},
  \citenamefont {{Schau{\ss}}}, \citenamefont {{Kondov}}, \citenamefont
  {{Khatami}}, \citenamefont {{Paiva}}, \citenamefont {{Trivedi}},
  \citenamefont {{Huse}},\ and\ \citenamefont {{Bakr}}}]{brown2016xx}%
  \BibitemOpen
  \bibfield  {author} {\bibinfo {author} {\bibfnamefont {P.~T.}\ \bibnamefont
  {{Brown}}}, \bibinfo {author} {\bibfnamefont {D.}~\bibnamefont {{Mitra}}},
  \bibinfo {author} {\bibfnamefont {E.}~\bibnamefont {{Guardado-Sanchez}}},
  \bibinfo {author} {\bibfnamefont {P.}~\bibnamefont {{Schau{\ss}}}}, \bibinfo
  {author} {\bibfnamefont {S.~S.}\ \bibnamefont {{Kondov}}}, \bibinfo {author}
  {\bibfnamefont {E.}~\bibnamefont {{Khatami}}}, \bibinfo {author}
  {\bibfnamefont {T.}~\bibnamefont {{Paiva}}}, \bibinfo {author} {\bibfnamefont
  {N.}~\bibnamefont {{Trivedi}}}, \bibinfo {author} {\bibfnamefont {D.~A.}\
  \bibnamefont {{Huse}}}, \ and\ \bibinfo {author} {\bibfnamefont {W.~S.}\
  \bibnamefont {{Bakr}}},\ }\href@noop {} {\  (\bibinfo {year} {2016})},\
  \Eprint {http://arxiv.org/abs/1612.07746} {arXiv:1612.07746
  [cond-mat.quant-gas]} \BibitemShut {NoStop}%
\bibitem [{\citenamefont {Boll}\ \emph {et~al.}(2016)\citenamefont {Boll},
  \citenamefont {Hilker}, \citenamefont {Salomon}, \citenamefont {Omran},
  \citenamefont {Nespolo}, \citenamefont {Pollet}, \citenamefont {Bloch},\ and\
  \citenamefont {Gross}}]{boll2016}%
  \BibitemOpen
  \bibfield  {author} {\bibinfo {author} {\bibfnamefont {M.}~\bibnamefont
  {Boll}}, \bibinfo {author} {\bibfnamefont {T.~A.}\ \bibnamefont {Hilker}},
  \bibinfo {author} {\bibfnamefont {G.}~\bibnamefont {Salomon}}, \bibinfo
  {author} {\bibfnamefont {A.}~\bibnamefont {Omran}}, \bibinfo {author}
  {\bibfnamefont {J.}~\bibnamefont {Nespolo}}, \bibinfo {author} {\bibfnamefont
  {L.}~\bibnamefont {Pollet}}, \bibinfo {author} {\bibfnamefont
  {I.}~\bibnamefont {Bloch}}, \ and\ \bibinfo {author} {\bibfnamefont
  {C.}~\bibnamefont {Gross}},\ }\href {\doibase 10.1126/science.aag1635}
  {\bibfield  {journal} {\bibinfo  {journal} {Science}\ }\textbf {\bibinfo
  {volume} {353}},\ \bibinfo {pages} {1257} (\bibinfo {year}
  {2016})}\BibitemShut {NoStop}%
\bibitem [{\citenamefont {Endres}\ \emph {et~al.}(2011)\citenamefont {Endres},
  \citenamefont {Cheneau}, \citenamefont {Fukuhara}, \citenamefont
  {Weitenberg}, \citenamefont {Schauß}, \citenamefont {Gross}, \citenamefont
  {Mazza}, \citenamefont {Bañuls}, \citenamefont {Pollet}, \citenamefont
  {Bloch},\ and\ \citenamefont {Kuhr}}]{endres2011}%
  \BibitemOpen
  \bibfield  {author} {\bibinfo {author} {\bibfnamefont {M.}~\bibnamefont
  {Endres}}, \bibinfo {author} {\bibfnamefont {M.}~\bibnamefont {Cheneau}},
  \bibinfo {author} {\bibfnamefont {T.}~\bibnamefont {Fukuhara}}, \bibinfo
  {author} {\bibfnamefont {C.}~\bibnamefont {Weitenberg}}, \bibinfo {author}
  {\bibfnamefont {P.}~\bibnamefont {Schauß}}, \bibinfo {author} {\bibfnamefont
  {C.}~\bibnamefont {Gross}}, \bibinfo {author} {\bibfnamefont
  {L.}~\bibnamefont {Mazza}}, \bibinfo {author} {\bibfnamefont {M.~C.}\
  \bibnamefont {Bañuls}}, \bibinfo {author} {\bibfnamefont {L.}~\bibnamefont
  {Pollet}}, \bibinfo {author} {\bibfnamefont {I.}~\bibnamefont {Bloch}}, \
  and\ \bibinfo {author} {\bibfnamefont {S.}~\bibnamefont {Kuhr}},\ }\href
  {\doibase 10.1126/science.1209284} {\bibfield  {journal} {\bibinfo  {journal}
  {Science}\ }\textbf {\bibinfo {volume} {334}},\ \bibinfo {pages} {200}
  (\bibinfo {year} {2011})}\BibitemShut {NoStop}%
\bibitem [{som()}]{som}%
  \BibitemOpen
  \href@noop {} {\ }\bibinfo {note} {{See supporting online
  material}}\BibitemShut {NoStop}%
\bibitem [{\citenamefont {Woynarovich}(1982)}]{woynarovich1982}%
  \BibitemOpen
  \bibfield  {author} {\bibinfo {author} {\bibfnamefont {F.}~\bibnamefont
  {Woynarovich}},\ }\href {\doibase 10.1088/0022-3719/15/1/007} {\bibfield
  {journal} {\bibinfo  {journal} {J. Phys. C Solid State}\ }\textbf {\bibinfo
  {volume} {15}},\ \bibinfo {pages} {85} (\bibinfo {year} {1982})}\BibitemShut
  {NoStop}%
\bibitem [{\citenamefont {Imambekov}\ \emph {et~al.}(2012)\citenamefont
  {Imambekov}, \citenamefont {Schmidt},\ and\ \citenamefont
  {Glazman}}]{imambekov2012}%
  \BibitemOpen
  \bibfield  {author} {\bibinfo {author} {\bibfnamefont {A.}~\bibnamefont
  {Imambekov}}, \bibinfo {author} {\bibfnamefont {T.~L.}\ \bibnamefont
  {Schmidt}}, \ and\ \bibinfo {author} {\bibfnamefont {L.~I.}\ \bibnamefont
  {Glazman}},\ }\href {\doibase 10.1103/RevModPhys.84.1253} {\bibfield
  {journal} {\bibinfo  {journal} {Rev. Mod. Phys.}\ }\textbf {\bibinfo {volume}
  {84}},\ \bibinfo {pages} {1253} (\bibinfo {year} {2012})}\BibitemShut
  {NoStop}%
\bibitem [{\citenamefont {Zaanen}(1998)}]{zaanen1998}%
  \BibitemOpen
  \bibfield  {author} {\bibinfo {author} {\bibfnamefont {J.}~\bibnamefont
  {Zaanen}},\ }\href {\doibase http://dx.doi.org/10.1016/S0022-3697(98)00106-1}
  {\bibfield  {journal} {\bibinfo  {journal} {Journal of Physics and Chemistry
  of Solids}\ }\textbf {\bibinfo {volume} {59}},\ \bibinfo {pages} {1769 }
  (\bibinfo {year} {1998})}\BibitemShut {NoStop}%
\bibitem [{\citenamefont {Recati}\ \emph {et~al.}(2003)\citenamefont {Recati},
  \citenamefont {Fedichev}, \citenamefont {Zwerger},\ and\ \citenamefont
  {Zoller}}]{recati2003}%
  \BibitemOpen
  \bibfield  {author} {\bibinfo {author} {\bibfnamefont {A.}~\bibnamefont
  {Recati}}, \bibinfo {author} {\bibfnamefont {P.~O.}\ \bibnamefont
  {Fedichev}}, \bibinfo {author} {\bibfnamefont {W.}~\bibnamefont {Zwerger}}, \
  and\ \bibinfo {author} {\bibfnamefont {P.}~\bibnamefont {Zoller}},\ }\href
  {\doibase 10.1103/PhysRevLett.90.020401} {\bibfield  {journal} {\bibinfo
  {journal} {Phys. Rev. Lett.}\ }\textbf {\bibinfo {volume} {90}},\ \bibinfo
  {pages} {020401} (\bibinfo {year} {2003})}\BibitemShut {NoStop}%
\bibitem [{\citenamefont {Kollath}\ \emph {et~al.}(2005)\citenamefont
  {Kollath}, \citenamefont {Schollw\"ock},\ and\ \citenamefont
  {Zwerger}}]{kollath2005}%
  \BibitemOpen
  \bibfield  {author} {\bibinfo {author} {\bibfnamefont {C.}~\bibnamefont
  {Kollath}}, \bibinfo {author} {\bibfnamefont {U.}~\bibnamefont
  {Schollw\"ock}}, \ and\ \bibinfo {author} {\bibfnamefont {W.}~\bibnamefont
  {Zwerger}},\ }\href {\doibase 10.1103/PhysRevLett.95.176401} {\bibfield
  {journal} {\bibinfo  {journal} {Phys. Rev. Lett.}\ }\textbf {\bibinfo
  {volume} {95}},\ \bibinfo {pages} {176401} (\bibinfo {year}
  {2005})}\BibitemShut {NoStop}%
\bibitem [{\citenamefont {Carlson}\ \emph {et~al.}(2004)\citenamefont
  {Carlson}, \citenamefont {Kivelson}, \citenamefont {Orgad},\ and\
  \citenamefont {Emery}}]{carlson2004}%
  \BibitemOpen
  \bibfield  {author} {\bibinfo {author} {\bibfnamefont {E.~W.}\ \bibnamefont
  {Carlson}}, \bibinfo {author} {\bibfnamefont {S.~A.}\ \bibnamefont
  {Kivelson}}, \bibinfo {author} {\bibfnamefont {D.}~\bibnamefont {Orgad}}, \
  and\ \bibinfo {author} {\bibfnamefont {V.~J.}\ \bibnamefont {Emery}},\
  }\enquote {\bibinfo {title} {Concepts in high temperature
  superconductivity},}\ in\ \href {\doibase 10.1007/978-3-642-18914-2_6} {\emph
  {\bibinfo {booktitle} {The Physics of Superconductors: Vol. II.
  Superconductivity in Nanostructures, High-T c and Novel Superconductors,
  Organic Superconductors}}},\ \bibinfo {editor} {edited by\ \bibinfo {editor}
  {\bibfnamefont {K.~H.}\ \bibnamefont {Bennemann}}\ and\ \bibinfo {editor}
  {\bibfnamefont {J.~B.}\ \bibnamefont {Ketterson}}}\ (\bibinfo  {publisher}
  {Springer Berlin Heidelberg},\ \bibinfo {address} {Berlin, Heidelberg},\
  \bibinfo {year} {2004})\ pp.\ \bibinfo {pages} {275--451}\BibitemShut
  {NoStop}%
\bibitem [{\citenamefont {Zhang}\ \emph {et~al.}(2002)\citenamefont {Zhang},
  \citenamefont {Demler},\ and\ \citenamefont {Sachdev}}]{zhang2002}%
  \BibitemOpen
  \bibfield  {author} {\bibinfo {author} {\bibfnamefont {Y.}~\bibnamefont
  {Zhang}}, \bibinfo {author} {\bibfnamefont {E.}~\bibnamefont {Demler}}, \
  and\ \bibinfo {author} {\bibfnamefont {S.}~\bibnamefont {Sachdev}},\ }\href
  {\doibase 10.1103/PhysRevB.66.094501} {\bibfield  {journal} {\bibinfo
  {journal} {Phys. Rev. B}\ }\textbf {\bibinfo {volume} {66}},\ \bibinfo
  {pages} {094501} (\bibinfo {year} {2002})}\BibitemShut {NoStop}%
\bibitem [{\citenamefont {Senthil}\ \emph {et~al.}(2004)\citenamefont
  {Senthil}, \citenamefont {Vishwanath}, \citenamefont {Balents}, \citenamefont
  {Sachdev},\ and\ \citenamefont {Fisher}}]{senthil2004}%
  \BibitemOpen
  \bibfield  {author} {\bibinfo {author} {\bibfnamefont {T.}~\bibnamefont
  {Senthil}}, \bibinfo {author} {\bibfnamefont {A.}~\bibnamefont {Vishwanath}},
  \bibinfo {author} {\bibfnamefont {L.}~\bibnamefont {Balents}}, \bibinfo
  {author} {\bibfnamefont {S.}~\bibnamefont {Sachdev}}, \ and\ \bibinfo
  {author} {\bibfnamefont {M.~P.~A.}\ \bibnamefont {Fisher}},\ }\href {\doibase
  10.1126/science.1091806} {\bibfield  {journal} {\bibinfo  {journal}
  {Science}\ }\textbf {\bibinfo {volume} {303}},\ \bibinfo {pages} {1490}
  (\bibinfo {year} {2004})}\BibitemShut {NoStop}%
\bibitem [{\citenamefont {Wen}(2004)}]{wen2004}%
  \BibitemOpen
  \bibfield  {author} {\bibinfo {author} {\bibfnamefont {X.-G.}\ \bibnamefont
  {Wen}},\ }\href@noop {} {\emph {\bibinfo {title} {Quantum Field Theory of
  Many-Body Systems}}}\ (\bibinfo  {publisher} {{Oxford University Press}},\
  \bibinfo {year} {2004})\BibitemShut {NoStop}%
\bibitem [{\citenamefont {Z{\"u}rn}\ \emph {et~al.}(2013)\citenamefont
  {Z{\"u}rn}, \citenamefont {Lompe}, \citenamefont {Wenz}, \citenamefont
  {Jochim}, \citenamefont {Julienne},\ and\ \citenamefont {Hutson}}]{zurn2013}%
  \BibitemOpen
  \bibfield  {author} {\bibinfo {author} {\bibfnamefont {G.}~\bibnamefont
  {Z{\"u}rn}}, \bibinfo {author} {\bibfnamefont {T.}~\bibnamefont {Lompe}},
  \bibinfo {author} {\bibfnamefont {A.}~\bibnamefont {Wenz}}, \bibinfo {author}
  {\bibfnamefont {S.}~\bibnamefont {Jochim}}, \bibinfo {author} {\bibfnamefont
  {P.}~\bibnamefont {Julienne}}, \ and\ \bibinfo {author} {\bibfnamefont
  {J.}~\bibnamefont {Hutson}},\ }\href {\doibase
  10.1103/physrevlett.110.135301} {\bibfield  {journal} {\bibinfo  {journal}
  {Phys. Rev. Lett.}\ }\textbf {\bibinfo {volume} {110}},\ \bibinfo {pages}
  {135301} (\bibinfo {year} {2013})}\BibitemShut {NoStop}%
\bibitem [{\citenamefont {B\"uchler}(2010)}]{buchler2010}%
  \BibitemOpen
  \bibfield  {author} {\bibinfo {author} {\bibfnamefont {H.~P.}\ \bibnamefont
  {B\"uchler}},\ }\href {\doibase 10.1103/PhysRevLett.104.090402} {\bibfield
  {journal} {\bibinfo  {journal} {Phys. Rev. Lett.}\ }\textbf {\bibinfo
  {volume} {104}},\ \bibinfo {pages} {090402} (\bibinfo {year}
  {2010})}\BibitemShut {NoStop}%
\bibitem [{\citenamefont {Serwane}\ \emph {et~al.}(2011)\citenamefont
  {Serwane}, \citenamefont {Z{\"u}rn}, \citenamefont {Lompe}, \citenamefont
  {Ottenstein}, \citenamefont {Wenz},\ and\ \citenamefont
  {Jochim}}]{serwane2011}%
  \BibitemOpen
  \bibfield  {author} {\bibinfo {author} {\bibfnamefont {F.}~\bibnamefont
  {Serwane}}, \bibinfo {author} {\bibfnamefont {G.}~\bibnamefont {Z{\"u}rn}},
  \bibinfo {author} {\bibfnamefont {T.}~\bibnamefont {Lompe}}, \bibinfo
  {author} {\bibfnamefont {T.~B.}\ \bibnamefont {Ottenstein}}, \bibinfo
  {author} {\bibfnamefont {A.~N.}\ \bibnamefont {Wenz}}, \ and\ \bibinfo
  {author} {\bibfnamefont {S.}~\bibnamefont {Jochim}},\ }\href {\doibase
  10.1126/science.1201351} {\bibfield  {journal} {\bibinfo  {journal}
  {Science}\ }\textbf {\bibinfo {volume} {332}},\ \bibinfo {pages} {336}
  (\bibinfo {year} {2011})}\BibitemShut {NoStop}%
\bibitem [{\citenamefont {Rigol}\ \emph {et~al.}(2003)\citenamefont {Rigol},
  \citenamefont {Muramatsu}, \citenamefont {Batrouni},\ and\ \citenamefont
  {Scalettar}}]{rigol2003}%
  \BibitemOpen
  \bibfield  {author} {\bibinfo {author} {\bibfnamefont {M.}~\bibnamefont
  {Rigol}}, \bibinfo {author} {\bibfnamefont {A.}~\bibnamefont {Muramatsu}},
  \bibinfo {author} {\bibfnamefont {G.~G.}\ \bibnamefont {Batrouni}}, \ and\
  \bibinfo {author} {\bibfnamefont {R.~T.}\ \bibnamefont {Scalettar}},\ }\href
  {\doibase 10.1103/PhysRevLett.91.130403} {\bibfield  {journal} {\bibinfo
  {journal} {Phys. Rev. Lett.}\ }\textbf {\bibinfo {volume} {91}},\ \bibinfo
  {pages} {130403} (\bibinfo {year} {2003})}\BibitemShut {NoStop}%
\bibitem [{\citenamefont {Coll}(1974)}]{Coll1974}%
  \BibitemOpen
  \bibfield  {author} {\bibinfo {author} {\bibfnamefont {C.~F.}\ \bibnamefont
  {Coll}},\ }\href {\doibase 10.1103/PhysRevB.9.2150} {\bibfield  {journal}
  {\bibinfo  {journal} {Phys. Rev. B}\ }\textbf {\bibinfo {volume} {9}},\
  \bibinfo {pages} {2150} (\bibinfo {year} {1974})}\BibitemShut {NoStop}%
\bibitem [{\citenamefont {Aichhorn}\ \emph {et~al.}(2003)\citenamefont
  {Aichhorn}, \citenamefont {Daghofer}, \citenamefont {Evertz},\ and\
  \citenamefont {von~der Linden}}]{Aichhorn2003}%
  \BibitemOpen
  \bibfield  {author} {\bibinfo {author} {\bibfnamefont {M.}~\bibnamefont
  {Aichhorn}}, \bibinfo {author} {\bibfnamefont {M.}~\bibnamefont {Daghofer}},
  \bibinfo {author} {\bibfnamefont {H.~G.}\ \bibnamefont {Evertz}}, \ and\
  \bibinfo {author} {\bibfnamefont {W.}~\bibnamefont {von~der Linden}},\ }\href
  {\doibase 10.1103/PhysRevB.67.161103} {\bibfield  {journal} {\bibinfo
  {journal} {Phys. Rev. B}\ }\textbf {\bibinfo {volume} {67}},\ \bibinfo
  {pages} {161103} (\bibinfo {year} {2003})}\BibitemShut {NoStop}%
\end{thebibliography}%
\section*{Supplementary material}

\renewcommand{\l}{\left(}
\renewcommand{\r}{\right)}
\newcommand{\T}{\mathcal{T}}
\newcommand{\gs}{\text{gs}}

\newcommand{\bra}[1]{\langle#1|}
\newcommand{\bkt}[2]{\left\langle #1 |#2 \right\rangle}
\renewcommand{\ij}{{\langle i, j \rangle}}
\renewcommand{\H}{\hat{\mathcal{H}}}
\newcommand{\Ht}{\tilde{\mathcal{H}}}
\renewcommand{\c}{\hat{c}}
\renewcommand{\a}{\hat{a}}
\newcommand{\cd}{\hat{c}^\dagger}
\newcommand{\rh}{\hat{\rho}}
\newcommand{\rht}{\tilde{\rho}}
\newcommand{\ad}{\hat{a}^\dagger}
\newcommand{\bd}{\hat{b}^\dagger}
\newcommand{\ubd}{\hat{\uline{b}}^\dagger}
\newcommand{\ub}{\hat{\uline{b}}}
\renewcommand{\b}{\hat{b}}
\newcommand{\hd}{\hat{h}^\dagger}
\newcommand{\h}{\hat{h}}
\newcommand{\dd}{\hat{d}^\dagger}
\renewcommand{\d}{\hat{d}}
\newcommand{\n}{\hat{n}}
\newcommand{\D}{\hat{D}}
\newcommand{\Dd}{\hat{D}^\dagger~\hspace{-0.12cm}}

\newcommand{\G}{\hat{\Gamma}}
\newcommand{\Gd}{\hat{\Gamma}^\dagger}
\newcommand{\F}{\hat{F}}
\newcommand{\Fd}{\hat{F}^\dagger}
\newcommand{\hc}{\text{h.c.}}
\newcommand{\MF}{\text{MF}}
\newcommand{\BEC}{\text{BEC}}
\newcommand{\RG}{\text{RG}}
\newcommand{\psd}{\hat{\psi}^\dagger}
\newcommand{\ps}{\hat{\psi}}
\newcommand{\I}{\text{I}}
\newcommand{\p}{\text{p}}
\newcommand{\f}{\text{F}}
\newcommand{\s}{\text{S}}
\renewcommand{\sf}{\text{MIX}}
\renewcommand{\O}{\hat{\mathcal{O}}}
\newcommand{\U}{\hat{U}}
\newcommand{\W}{\hat{W}}
\newcommand{\Ud}{\hat{U}^\dagger}
\newcommand{\KP}{\text{KP}}
\newcommand{\HMF}{\mathscr{H}_{\text{MF}}}
\newcommand{\ph}{\text{ph}}
\newcommand{\IB}{\text{IB}}
\newcommand{\B}{\text{B}}
\newcommand{\eff}{\text{eff}}
\newcommand{\tr}{\text{tr}}
\newcommand{\comm}[1]{{\color{black} {#1}}}

\newcommand{\blankpage}{
\newpage
\thispagestyle{empty}
\mbox{}
\newpage
}



\setcounter{figure}{0}
\setcounter{equation}{0}
\renewcommand{\thetable}{S\arabic{table}}
\renewcommand{\thefigure}{S\arabic{figure}}
\renewcommand{\theequation}{S\arabic{equation}}%



\section{Preparation of the ultracold lattice gas}

The preparation of degenerate Hubbard chains closely followed the protocol
detailed in Ref.~\cite{boll2016}.  We started from a quasi two-dimensional
degenerate mixture of the two lowest energy Zeeman states of $^6$Li in a single
plane of an optical lattice with $3.1\,\mu$m spacing in the vertical direction.
The vertical lattice depth was $110\,E_r^z$ and the scattering length was set
to $530\,a_B$. Here $E_r^i =h^2/8ma_i^2$ is the recoil energy for a lattice of
period $a_i$ in the $i$-direction, $m$ the atomic mass and $a_B$ the Bohr
radius. Next, the preparation of about $10$ Fermi-Hubbard chains started with
ramping up the large scale component of an optical superlattice
($a_{sl}=2.3\,\mu$m) in the $y$-direction in $15\,$ms to a depth of
$18\,E_r^y$. A lattice of period $a_l=1.15\,\mu$m along $x$ was  then ramped up
in $15\,$ms to $3\,E_r^x$ and finally to $5\,E_r^x$ in $80\,$ms, while the
lattice depth in $z$-direction was linearly decreased to $17\,E_r^z$ in
$50\,$ms. Simultaneously the lattice in $y$-direction was increased to
$27\,E_r^y$ in $60\,$ms. Using a magnetic offset field of $714\,$G near the
broad Feshbach resonance located at $834.1\,$G \cite{zurn2013} the scattering
length was linearly increased during these ramps to $2000\,a_B$. The lattice
and the low peak densities of about one atom per site ensured collisional
stability by suppressing three-body recombination losses. At the end of the
ramps, the onsite interaction energy was $U=h\times2.9\,$kHz, as estimated
from Wannier function calculations without taking into account finite band gap
corrections~\cite{buchler2010}. The tunneling amplitude was $t=h\times 400\,$Hz
and the exchange energy $J=4t^2/U=h\times 220\,$Hz. A local Stern-Gerlach
detection technique operating at a transverse magnetic field gradient of
$95\,$G/cm detailed in~\cite{boll2016} was used to detect both the spin and the
density on each lattice site with a fidelity larger than $98\%$.

\section{Properties of the atomic cloud}
\begin{SCfigure*}[\sidecaptionrelwidth][ht]
	\centering
	\includegraphics[]{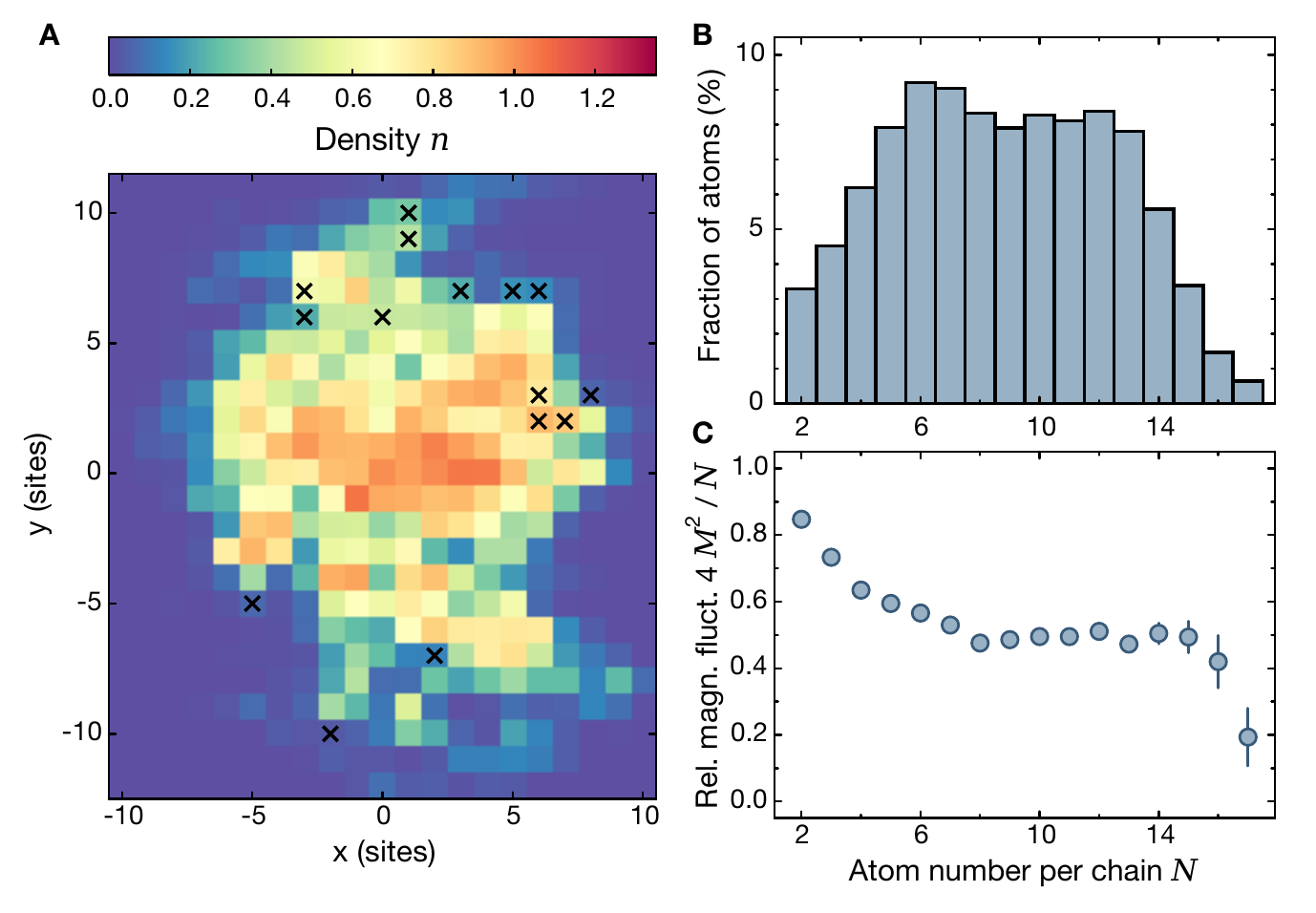}
\caption{
\textbf{Cloud properties} \textbf{(A)} Density distribution of the cloud
obtained by averaging $2700$ experimental runs. The decoupled chains run along
the $x$-direction. Each pixel corresponds to a lattice site and black crosses
identify sites which were filtered out because spin detection failed due to
optical potential imperfections. \textbf{(B)} Distribution of the atoms into
chains of different atom number. The typical accessible chain length is $15$
sites, leading to a broad distribution of different densities in every shot.
\textbf{(C)} Magnetization fluctuations $M^2$ per chain normalized to the
expected value for uncorrelated fluctuations $N/4$. }
\end{SCfigure*}

The prepared clouds contained in total $131\pm5.5$ atoms with a
global magnetization of $\frac12 (N_{\uparrow}+N_{\downarrow})=+1.2\pm2.9$ compatible with zero, where the uncertainties are the standard
deviations of the distributions, i.e. not the standard error of the mean.
The sub-shot-noise fluctuations are attributed to our magnetic gradient
assisted evaporation in a stiff optical trap, which cuts into the Fermi
sea~\cite{serwane2011}. The gaussian intensity profiles of the lattices beams
introduced an additional confining potential, which lead to an inhomogeneous
density distribution (Fig.~S1~A). The atom number was chosen to obtain
$N\lesssim N_0= 2\left(\frac{4t}{\frac12 m\bar{\omega}^2a_{l}^2}\right)^{1/2}\simeq 13$
in most chains, corresponding to a filling lower than one atom per site~\cite{rigol2003}.
Here, the frequency corresponding to the harmonically approximated  confinement
along the chains is $\bar{\omega}\approx 2\pi \times 300\,$Hz.  The atom number
distribution in the tubes is shown in Fig.~S1~B.

Equipped with the measurement of the full counting statistics, we binned the
$38000$ chains from all shots of the dataset by their number of atoms $N = \langle\sum_i
(\hat{n}_{i\uparrow} + \hat{n}_{i\downarrow})\rangle$, where $i$ is running
over all sites of the chain. For each bin of fixed 
$N$, we then analyzed the magnetization $\hat{M}= \sum_i\hat{S}^z_i$
and magnetization fluctuations $\langle \hat{M}^2 \rangle$. For typical atom
numbers of $N=7-15$, we observed less than half a spin of net magnetization
$\langle \hat{M} \rangle = 0.011(1)\,N$ and sub-shot-noise fluctuations
$\langle 4\hat{M}^2/N \rangle = 0.519(6)$ per chain,  which is about half the value
expected for uncorrelated spins (see Fig.~S1~C). These fluctuations were taken
into account when comparing to numerical predictions.

\section{Data post selection}

The fidelity of the spin resolved imaging depends on the phase fluctuations of
the superlattice in $y$-direction. In $3\%$ of the experimental runs we
detected unusually large global spin imbalances, which we attribute to
environmentally induced lattice phase fluctuations. Thus, runs with more than
$16$ excess spins in the entire cloud have been discarded. The presence of
short scale imperfections in the trapping potential, visible in Fig.~S1A,
resulted in a consistent failure of spin detection on some
sites~\cite{boll2016}. While the precise superlattice phase control ensured a local
mean spin of $S^z_i = 0.004(3)$ for typical "good" sites, we filtered out
sites with detected mean spin of more than $S^z_i = 0.025$ on a $3.5\sigma$ level. This concerns $15$ lattice sites out
of about $300$ (Fig.~S1A). In about $1\%$ of the chains we detected one or more
site occupied by two atoms of the same spin. These events might be due to a small
chance of having atoms in the second band or due to hopping during the
detection. We removed all of those lines from the dataset. We ensured that none
of the filters discussed above critically affect the results reported in this
manuscript.


\section{Correlation functions}

\begin{figure*}[t!]
	\centering
	 \includegraphics[]{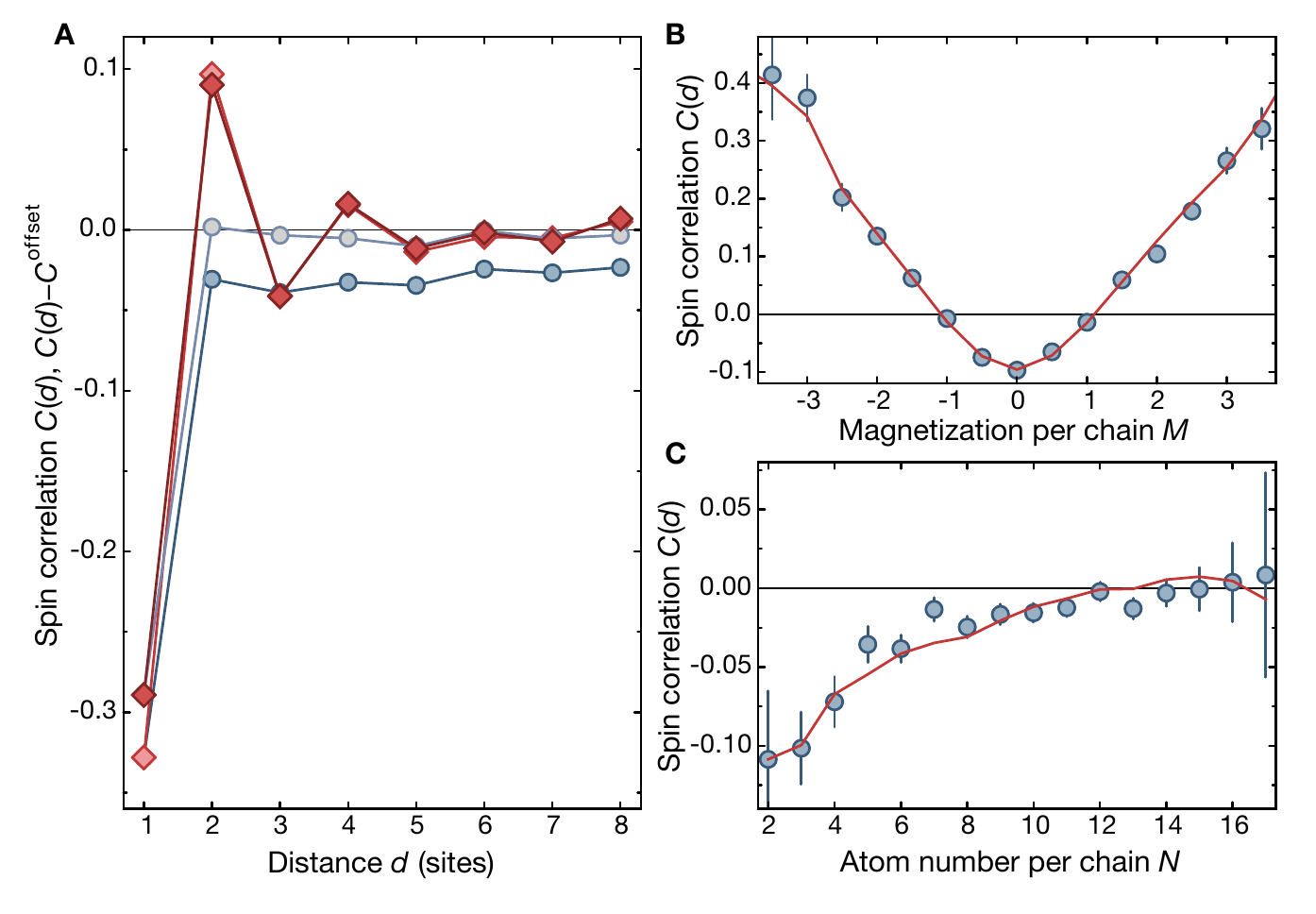}
\caption{
\textbf{Finite size offset analysis}
\textbf{(A)} Effect of offset subtraction. The spin correlation function $C(d)$
(dark blue) shows a clear offset that is expected in finite size systems with
squeezed magnetization fluctuations. Subtraction of this offset leads to a
correlation function which decays to zero (light blue). The spin string
correlation function, $C^{\textrm{str}}(d)$ (red), is almost unaffected (light red) because
of the changing signs $(-1)^{N_h}$ in the definition. Measured correlation
values for large distances $C(d>6)$ (blue points) as a function of \textbf{(B)}
chain magnetization $M$ and \textbf{(C)} atom number $N$. The estimated offset
corrections $C^{\text{offset}}(M,N)$ (red lines) from Eq.~\ref{Offset} with $A =
-0.045$ averaged over experimental atom number (B) or magnetization (C) distribution captures the experimental offset well.}
\end{figure*}

An empty or doubly occupied site has spin zero and thus these sites reduce
the magnitude of spin correlations trivially. In order to compare spin
correlations at different densities, we evaluate the spin operators on singly
occupied sites only. The conventional connected spin correlation function
including this condition can be written as
\begin{align*}
C(d) &= 4\,\langle\hat{S}^z_{i} \hat{S}^z_{i+d}\rangle_{\tiny\newmoon_i \newmoon_{i+d}}
\equiv 4 \frac{\langle\hat{S}^z_{i} \hat{S}^z_{i+d}\rangle}{\langle\hat{n}^s_{i} \hat{n}^s_{i+d} \rangle} 
\approx \frac4{n_i n_{i+d}}\langle\hat{S}^z_{i} \hat{S}^z_{i+d}\rangle,
\end{align*}
where we defined the "singlon" operator $\hat{n}_i^s =
\hat{n}_{i,\uparrow}+\hat{n}_{i,\downarrow}-2\hat{n}_{i,\uparrow}\hat{n}_{i,\downarrow}$.
The first relation is the definition of the conditional correlation function
$C(d)$, which we directly measured experimentally. The doublon fraction as well
as density correlations beyond one site were negligible, which justifies the
approximation of the "singlon-singlon" correlations in the denominator by the measured densities resulting 
in a simple normalized spin-spin correlator. These last expressions are just given for clarity, they have not been used in the analysis.

In the same spirit, the spin-hole correlation function, which selects on two
occupied and one empty site, can be rewritten as a normalized three-point
correlator:
\begin{widetext}

\begin{align*}
C_{SH}(d,s) &= 4\,\langle\hat{S}^z_{i} \hat{S}^z_{i+d} \rangle_{\tiny\newmoon_i \fullmoon_{i+s} \newmoon_{i+d}}
\equiv 4 \frac{\langle\hat{S}^z_{i} \hat{n}^h_{i+s} \hat{S}^z_{i+d}\rangle}{\langle\hat{n}^s_{i} \hat{n}^h_{i+s} \hat{n}^s_{i+d} \rangle} 
\approx \frac4{n_i (1-n_{i+s}) n_{i+d}}\langle\hat{S}^z_{i} \hat{n}^h_{i+s} \hat{S}^z_{i+d}\rangle,
\end{align*}
where the hole operator $\hat{n}_i^h =
(1-\hat{n}_{i,\uparrow})(1-\hat{n}_{i,\downarrow})\approx (1-n_{i})$ detects
the presence of a hole at site $i$.

The measured non-local correlation functions can equivalently be
expressed in terms of unconditional spin and density correlation functions,
\begin{align*}
C^{\textrm{str}}(d)&= 4\left\langle \hat{S}^z_i \left(\prod_{j=1}^{d-1}(-1)^{(1-\hat{n}_{i+j})}\right)
\hat{S}^z_{i+d}\right\rangle_{\tiny\newmoon_i\newmoon_{i+d}}
\equiv
 \frac4{\langle\hat{n}_i^s\hat{n}_{i+d}^s\rangle}\left\langle \hat{S}^z_i \left(\prod_{j=1}^{d-1}(-1)^{(1-\hat{n}_{i+j})}\right)
\hat{S}^z_{i+d}\right\rangle\\
C_{SH}^{\textrm{str}}(d,s)&= 4 \left\langle
\hat{S}^z_i  \left(\prod_{\substack{j=1,j\neq
s}}^{d-1}(-1)^{(1-\hat{n}_{i+j})}\right) \hat{S}^z_{i+d} \right\rangle_{\tiny\newmoon_i
\fullmoon_{i+s} \newmoon_{i+d}} 
\equiv
\frac4{\langle\hat{n}^s_{i} \hat{n}^h_{i+s} \hat{n}^s_{i+d} \rangle}  \left\langle \hat{S}^z_i \hat{n}_{i+s}^h  \left(\prod_{\substack{j=1,j\neq
s}}^{d-1}(-1)^{(1-\hat{n}_{i+j})}\right) \hat{S}^z_{i+d} \right\rangle.
\end{align*}
\end{widetext}


When studying both local and non-local spin correlations at large distances,
the presence of the trap complicates the definition of density. For the data
presented in the main text in Fig.~3 we defined the density $n$ as the mean
density over all the sites connecting the two operators evaluated at sites $i$
and $i+d$: $n =\frac{1}{d+1} \sum_{k=i}^{i+d}  \langle \hat{n}_k \rangle$.

\section{Correction for finite size effects}

\begin{figure}[t]
	\centering
	\includegraphics[width=\columnwidth]{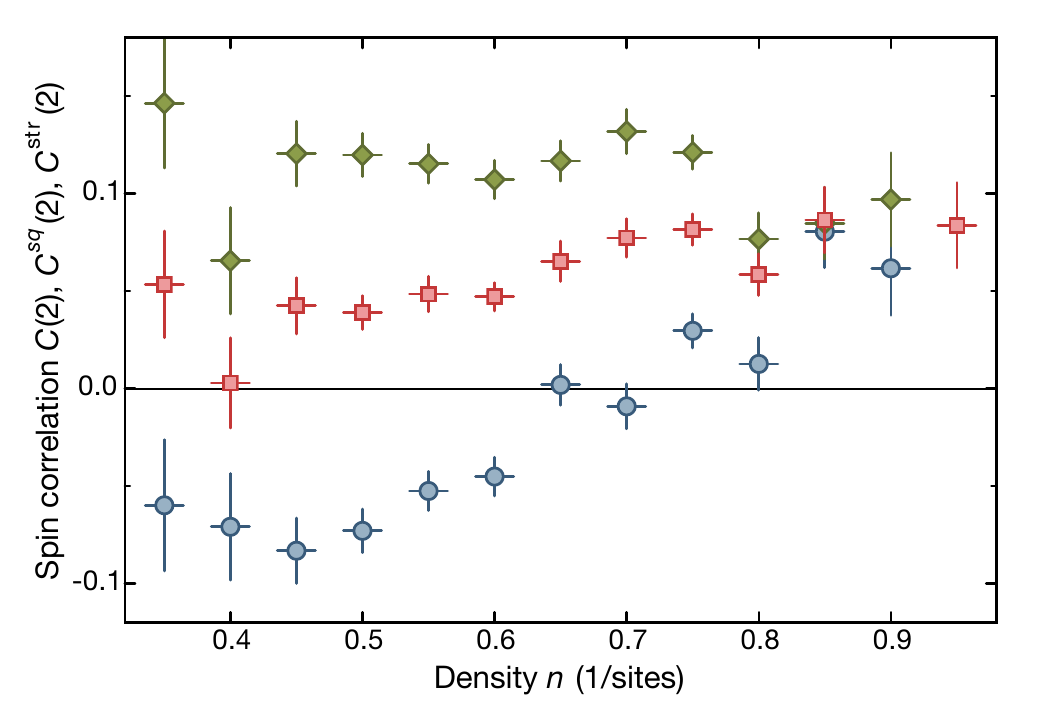}
\caption{
\textbf{Comparison of spin correlations.} Spin correlation $C(d=2)$ in blue,
spin string correlations $C^{\textrm{str}}(d=2)$ in green and squeezed space
correlations $C^{sq}(\tilde{d}=2)$ in red as a function of local density. 
}
\end{figure}

In each chain, the atom number and magnetization weakly fluctuate from
shot-to-shot, but for a single realization the atom number $N =\langle\sum_i
\hat{n}_i\rangle = N_{\uparrow} + N_{\downarrow}$ and magnetization $M=
\langle\sum_i \hat{S}^z_i\rangle =\frac12(N_{\uparrow} - N_{\downarrow})$ are
fixed. A fixed magnetization $M$ in a single spin-$1/2$ chain without holes
or doublons (length $L=N$) acts as a constraint on the spin-spin correlations because of the
following exact sum rule:

\begin{align}
4 \hat{M}^2 
		= \hat{N} +2 \sum_{d=1}^{N-1} (N-d) \hat{C}(d) \label{sumRule}
\end{align}
Here, the line average of the spin correlation operator $\hat{C}(d) =
4/(N-d)\sum_{i=1}^{N-d} \hat{S}_i^z \hat{S}_{i+d}^z$ has been introduced.\\
Even for completely uncorrelated spins in the chain the correlations are thus non-zero 
and their value $C(d) = 4 M^2 /(N^2-N) - 1/(N-1)$ is necessarily constant with distance $d$.
This effect only vanishes in the case of Poisson (shot-noise) magnetization
fluctuations $4 M^2=N$, or in the infinite system size limit assuming
non-extensive fluctuations. For sub-shot-noise fluctuations of the
magnetization this offset is negative, in agreement with our experimental and
numerical observations.\\
For systems with non-trivial correlations $C^{\textrm{corr}}$ we assume a
constant additive offset $C(d)=C^{\text{corr}}(d)+C^{\text{offset}}(N,M)$ due
to this finite size effect. Solving Eq.~\ref{sumRule}, reveals an offset
$C^{\textrm{offset}}(N,M)$ depending on the correlations:

 \begin{align}
C^{\text{offset}}(N,M)= &\frac{4 M^2 }{ N^2-N} - \frac1{N-1} \nonumber\\
 &-\underbrace{\sum_{d=1}^{N-1} \frac{2(N-d)}{N(N-1)} C^{\text{corr}}(d)}_{A}\label{Offset}
\end{align}

Experimentally, we find that our data is well described by
$C^{\textrm{offset}}(N,M)$ with constant $A = -0.045(5)$ obtained from the
correlations for $d>6$, where we do not observe any staggered correlations in the
conventional spin correlator (Fig.~S2 B,C).
In order to reveal the non-trivial staggered spin correlations, we measured the atom number and magnetization for each chain,
calculated $C^{\textrm{offset}}(N,M)$ using the experimental value for $A$, and
finally subtracted it from each single outcome contributing to
$\langle\hat{S}^z_{i} \hat{S}^z_{i+d}\rangle$ before averaging spatially and
over different experimental runs. This procedure was followed for all different
versions of correlation functions, even though the string correlators are
largely insensitive to the offset (Fig.~S2A).

\section{Comparison of correlation functions}

We analyzed three types of spin correlation functions, the conventional
connected two-point correlation function $C(d)$, the string correlation
function $C^{\textrm{str}}(d)$ and the squeezed space correlation function
$C^{sq}(\tilde{d})$. Without any holes all of these correlators give the same results. To
emphasize the difference between them at finite doping, Fig.~S3 shows an
exemplary direct comparison of the $C^{\rm x}(2)$ values as a function of density.
The conventional correlation function $C(d)$ is very sensitive to doping due
to the discussed AFM parity flips and $C(2)$ even changes sign at $n=0.70(3)$,
when the contribution from spin-hole-spin events $C_{SH,N_h=1}(d=2)$ dominates
over events without a hole $C_{SH,N_h=0}(d=2)$. The spin string correlation
function, on the other hand, is insensitive to the hole induced AFM parity
flips, thus, $C^{\textrm{str}}(2)$ stays positive and even increases slightly in
magnitude with decreasing density due to the effectively shorter spin-spin
distance in squeezed space. The squeezed space correlation function
$C^{sq}(\tilde{d}=2)$ is also insensitive to the AFM parity flips and stays positive, but decreases with doping at finite temperature due to the
decreasing effective coupling strength between the spins (see Sec.~\ref{secSM:SCsep}).


\section{Effective Heisenberg model in squeezed space}
\label{secSM:SCsep}

\begin{figure}[t!]
\centering
\epsfig{file=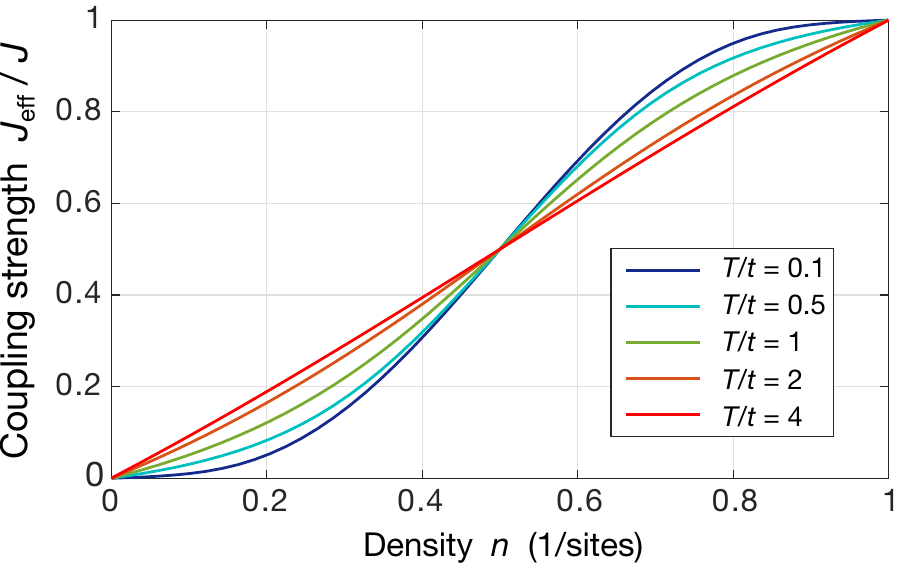,width=\columnwidth}
\caption{
\textbf{Effective spin interactions in squeezed space.} The effective spin exchange interaction $J_{\rm eff}$ as a function of density $n$, for different values of the temperature $T$ in the charge sector. For $T \gg t$ the holes are uncorrelated and $J_{\rm eff}$ scales linearly with the density. In the zero-temperature limit, on the other hand, $J_{\rm eff}$ depends on $n$ in a strongly non-linear way as a result of Pauli-blocking between neighboring holes.}
\label{fig:Jeff}
\end{figure}

In the limit of zero temperature and infinite repulsion $U/t \to \infty$, Ogata and Shiba have shown that the ground state wavefunction of the doped Fermi-Hubbard model factorizes \cite{ogata1990}, $\Psi(\{x_{j,\sigma}\}) =
\Psi_{\rm ch}(\{x_j\})\,\Psi_{\rm s}(\{\tilde{x}_{j,\sigma}\})$. The first part is a wavefunction of free, spinless fermions which describes the charge sector (holes). The second part corresponds to a pure spin wavefunction given by the ground state of an anti-ferromagnetic Heisenberg chain.  As described in the main text of our paper, this factorization can be illustrated in the squeezed space picture \cite{zaanen2001,kruis2004} where holes are removed from the lattice. 

For sufficiently large $U$, we expect that a similar factorization remains valid even for finite temperatures $T \geq 0$, i.e. spin and charge degrees of freedom remain uncorrelated. This is a direct consequence of the separation of the spin-exchange and hole-hopping energy scales, $J \ll t$. In analogy with Ogata and Shiba's wavefunction, we will make the following ansatz for describing the thermal density matrix $\hat{\rho}$ in the large-$U$ limit,
\begin{equation}
\hat{\rho} = \hat{\rho}_{\rm s} \otimes \hat{\rho}_{\rm ch}.
\label{eq:SCsepAnsatz}
\end{equation}

For the ground state at $T=0$, the presence of the holes has no effect on the wavefunction $\Psi_{\rm s}$ of spins on the squeezed lattice. Nevertheless the typical exchange energy scales are suppressed. Simply speaking, the spins are less likely to be located next to each other when the hole doping increases. At finite temperature $T>0$, in contrast to the ground state case, this reduction of the spin-exchange energies $J \to J_{\rm eff}$ modifies the spin-correlations. To describe this effect quantitatively we will now derive an expression for $J_{\rm eff}$, as a function of density $n$ and temperature $T$, see Eq.~\eqref{eq:Jeff} below. Our results are shown in Fig.~\ref{fig:Jeff}. Using this data we calculated the density dependence of spin-correlators presented in Fig.~3b of our paper.

\subsection{The $t-J^*$ model}
Our starting point is the Fermi-Hubbard model at fillings less than or equal to one half. We consider the limit when $U \gg t$, where double occupancy of sites by two fermions is strongly suppressed and the Fermi-Hubbard Hamiltonian reduces to the $t-J^*$ model. To leading order in $J=4 t^2/U$ one obtains the effective Hamiltonian
\begin{align}
\label{eq:HtJoriginal}
\H_{t-J*} = \hat{\mathcal{P}}_s &\left[ - t \sum_{i, j, \sigma} \cd_{i,\sigma} \c_{j,\sigma} \right. \\
&\ \left.+ J \sum_\ij \l \hat{\vec{S}}_i \cdot \hat{\vec{S}}_j - \frac{\n_i \n_j}{4} \r + \H_{\rm NNN} \right] \hat{\mathcal{P}}_s,\nonumber
\end{align}
see for example Ref.~\cite{auerbach1994}. Here $\c_{j,\sigma}$ annihilates a fermion with spin $\sigma$ on site $j$ and the spin operators are defined as $\hat{\vec{S}}_j = \frac{1}{2} \sum_{\sigma,\sigma'} \cd_{j,\sigma} \vec{\sigma}_{\sigma,\sigma'} \c_{j,\sigma'}$. We defined the fermion density as $\n_j = \sum_\sigma \cd_{\sigma,j} \c_{\sigma,j}$, and $\vec{\sigma}$ denotes a vector of Pauli matrices. Note that the Hamiltonian in Eq.~\eqref{eq:HtJoriginal} must be projected onto the subspace of single-occupied sites, which is ensured by the projection operators $\hat{\mathcal{P}}_s$. The first two terms in Eq.~\eqref{eq:HtJoriginal} correspond to the commonly studied $t-J$ Hamiltonian, where the additional next-nearest neighbor hopping processes described by $\H_{\rm NNN}$ are neglected. In our case such terms are crucial, however, and we will discuss them in more detail below.

The first term in Eq.~\eqref{eq:HtJoriginal} describes hopping processes of a fermion to an unoccupied neighboring site. The second term in Eq.~\eqref{eq:HtJoriginal} corresponds to the usual spin-exchange interaction between two fermions on neighboring sites. Its energy scale is set by $J=4 t^2/U$ because it derives from a second order tunneling processes where a state with two fermions on the same site is virtually occupied. The exchange interaction between two neighboring spins leads to a zero-point energy of $J/4$, which manifests itself by the nearest neighbor density-density interaction $J \n_i \n_j / 4$ in Eq.~\eqref{eq:HtJoriginal}.

\begin{figure}[b!]
\centering
\epsfig{file=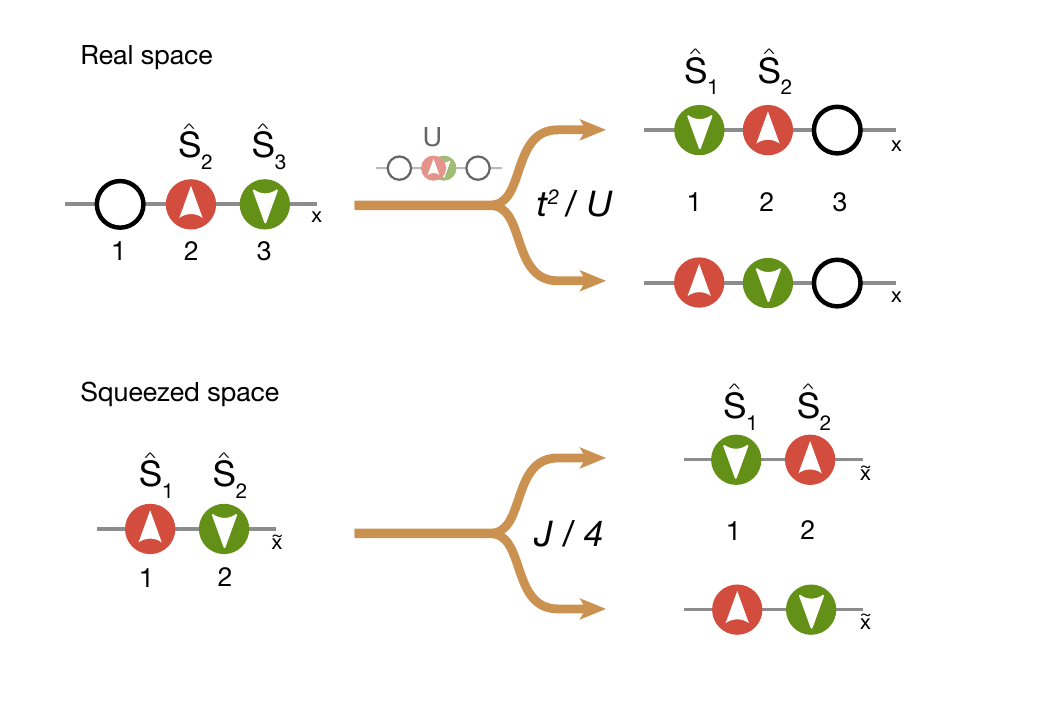, width=\columnwidth}
\caption{Illustration of the next-nearest neighbor hopping process in physical and squeezed space. The hopping of the hole is accompanied by a spin-exchange interaction $(\hat{\vec{S}}_{\tilde{j}} \cdot \hat{\vec{S}}_{\tilde{k}} - \frac{1}{4})$ in squeezed space.}
\label{fig:NNNhopping}
\end{figure}

In addition to the nearest neighbor spin-exchange interaction, virtual processes where a site is occupied by two fermions also leads to a next-nearest neighbor tunneling process of a hole. As illustrated in Fig.~\ref{fig:NNNhopping}, a hole can tunnel from site $i$ to $k=i+2$ in the following way. First a fermion hops from site $k=i+2$ onto another fermion on site $j=i+1$, where a virtual two-fermion state with energy $U$ is formed. Then one of the two fermions hops from site $j=i+1$ to site $i$. No double-occupancies are remaining now, and the hole has effectively moved to site $k=i+2$. The energy scale for this process is also set by $J$, and it can be formally described by the following term in the $t-J^*$ Hamiltonian~\cite{auerbach1994},
\begin{equation*}
\begin{split}
\H_{\rm NNN}  = &- \frac{J}{8}  \sum_{\langle i,j,k \rangle, \sigma}^{i \neq k}  \l \cd_{i,\sigma} \c_{k,\sigma} \n_j \vphantom{\sum_{\sigma', \tau, \tau'}} \right. \\
  &\left. - \sum_{\sigma', \tau, \tau'} \cd_{i,\sigma} \vec{\sigma}_{\sigma,\sigma'} \c_{k,\sigma'} \cdot  \cd_{j,\tau} \vec{\sigma}_{\tau, \tau'} \c_{j,\tau'} \r,
\end{split}
\end{equation*}

where $\langle i,j,k \rangle$ denotes a triple of neighboring sites.

\subsection{The $t-J^*$ model in squeezed space}
To understand the physics of the one-dimensional $t-J^*$ model \eqref{eq:HtJoriginal}, we would like to introduce spin-less hole operators describing the charge degrees of freedom as well as spin operators defined in squeezed space. First the hole operators $\h_j$ can be introduced by describing the spins with Schwinger bosons. This leads us to the following Hamiltonian (up to a constant energy offset), see Ref.~\cite{auerbach1994},
\begin{align}
\begin{split}
\label{eq:HtJ}
\H_{t-J*} ={}& t \sum_{\ij} \l \hd_i \h_j \hat{\mathcal{F}}_{ij}^\dagger + \hc \r \\
&+ J \sum_\ij  \l 1 - \n^h_i \r  \hat{\vec{S}}_i \cdot \hat{\vec{S}}_j \l 1 - \n^h_j \r  \\ 	
	&- \frac{J}{4} \sum_\ij \n^h_i \n^h_j + \frac{J}{4} \sum_{\langle i,j,k \rangle}^{i \neq k} \hd_k \l 1 - \n^h_j \r \h_i \hat{\mathcal{A}}_{ij}^\dagger \hat{\mathcal{A}}_{kj}.
\end{split}		
\end{align}
Here $\hd_i$ creates a hole on site $i$, $\n_i^h = \hd_i \h_i$ is the hole density, and $\hat{\vec{S}}_i$ denotes the spin operator at site $i$. The operators $\hat{\mathcal{F}}_{ij}$ and $\hat{\mathcal{A}}_{ij}$ can be expressed in terms of Schwinger bosons, see Ref.~\cite{auerbach1994} for details, but we will explain their meaning below by mapping the Hamiltonian to squeezed space.

The first term in Eq.~\eqref{eq:HtJ} describes the hopping of the holes, where the operators $\hat{\mathcal{F}}_{ij}$ ensure that the spins are physically moved in real space. This has no effect in squeezed space however, where the ordering of the spins is unchanged and we can effectively set $\hat{\mathcal{F}}_{ij}=1$. The second term describes the anti-ferromagnetic exchange interactions. Note that two neighboring spins in squeezed space can only interact if they are not separated by holes in real space. Thus, on average we expected a reduction of the exchange energy scale in squeezed space with increasing hole density.

The third term in Eq.~\eqref{eq:HtJ} describes an attractive interaction between the holes, which, however, is small compared to the kinetic energy term and can be neglected in the limit when $t \gg J$. 

The last term in Eq.~\eqref{eq:HtJ} describes the next-nearest neighbor tunneling of the hole from site $i$ to $k$, across a site $j$ which is occupied by a spin. This process is accompanied by a spin-flip interaction, described by the operators $\hat{\mathcal{A}}_{ij}$. The term can be simplified considerably by mapping it to a squeezed space representation: Consider a situation with two spins on sites $k=-1$ and $j=0$ and a hole on site $i=1$. In squeezed space we can label the two spins by $\tilde{j}$ and $\tilde{k}$, and their labels do not change when the hole hops from $i$ to $k$, see Fig.~\ref{fig:NNNhopping}. In the new basis, this next-nearest neighbor hopping process becomes
\begin{equation}
 \frac{J}{4}\hat{\mathcal{A}}_{ij}^\dagger \hat{\mathcal{A}}_{kj} = \frac{J}{2}  \l \hat{\vec{S}}_{\tilde{j}} \cdot \hat{\vec{S}}_{\tilde{k}} - \frac{1}{4} \r,
 \label{eq:NNNhopping}
\end{equation}
and the hole operators $\hd_k \l 1 - \n^h_j \r \h_i $ are unmodified.

Like before, in the limit $t \gg J$ the next-nearest neighbor hopping term $\sim J$ has no effect on the hole dynamics which is dominated by the nearest neighbor hopping term of order $\sim t$. However, the spin-dynamics can be substantially modified by this term, because it is of comparable strength $\sim J$ as the exchange interactions.

\subsection{Effective Hamiltonians: spin-charge separation}
Now we make use of the separation of energy scales, $t \gg J$ valid in the large-$U$ limit. This justifies neglecting spin-charge correlations and thus the use of our ansatz Eq.~\eqref{eq:SCsepAnsatz}. As a result we obtain an effective Hamiltonian for the holes,
\begin{equation}
\H_{\rm ch} = t \sum_{\ij} \l \hd_i \h_j + \hc \r = 2t  \sum_k \cos(k) \hd_k \h_k,
\label{eq:DispHoles}
\end{equation}
which is derived from the $t-J^*$ model Eq.~\eqref{eq:HtJ} by discarding terms of order $\mathcal{O}(J/t)$. As expected, this corresponds to free fermions hopping on a one-dimensional lattice. 

From the correlations of the holes, determined by $\hat{\rho}_{\rm ch}$, we can now derive an effective Hamiltonian for the spin sector. Formally it is obtained by tracing out the charge sector, i.e. $\H_{\rm s} = {\rm tr}_{\rm ch} ( \hat{\rho}_{\rm ch} \H_{t-J*} )$. As a result we obtain a pure spin Hamiltonian, 
\begin{equation*}
\H_{\rm s} = J_{\rm eff}(n) \sum_{\langle \tilde{i},\tilde{j} \rangle} \hat{\vec{S}}_{\tilde{i}} \cdot \hat{\vec{S}}_{\tilde{j}}.
\end{equation*}
Note that in contrast to $\hat{\vec{S}}_{j}$ in Eq.~\eqref{eq:HtJ}, $\hat{\vec{S}}_{\tilde{j}}$ denotes operators for the spin chain in squeezed space now.

\subsection{Effective spin-exchange interaction}
To obtain the effective exchange interaction $J_{\rm eff}$ in squeezed space, we need to calculate conditional probabilities in real space. To understand this, let us start by considering the second term in Eq.~\eqref{eq:HtJ}. If there is a spin on a given site $i$ in real space, corresponding to spin $r$ in squeezed space, it interacts with its neighboring spin $s=r+1$ in squeezed space only if there is a spin on site $i+1$ in real space. This leads to a contribution $J_{\rm eff}^{(1)}$ to the exchange interactions in squeezed space, given by
\begin{equation*}
J_{\rm eff}^{(1)} = J \frac{\left\langle \l 1-\n_i^h \r \l 1-\n_{i+1}^h \r \right\rangle}{\left\langle \l 1-\n_i^h \r \right\rangle}.
\end{equation*}

Similarly we obtain a contribution $J_{\rm eff}^{(2)}$ to the spin-exchange interactions from the next-nearest neighbor hopping processes from Eq.~\eqref{eq:NNNhopping}. In this case we need to calculate the following conditional correlator,
\begin{equation*}
J_{\rm eff}^{(2)} = \frac{J}{2} \frac{\left\langle \l 1-\n_i^h \r \l \hd_{i-1} \h_{i} +  \hd_{i} \h_{i-1} \r \right\rangle}{\left\langle \l 1-\n_i^h \r \right\rangle}.
\end{equation*}

Assuming that the holes at density $n_h=1-n$ have a temperature $T$, we can easily calculate all correlation functions assuming a Fermi-Dirac distribution function $n^F_k(n,T)$ of the holes (Eq.~\eqref{eq:DispHoles}) and making use of Wick's theorem. As a result we obtain for $J_{\rm eff} = J_{\rm eff}^{(1)}  + J_{\rm eff}^{(2)}$,
\begin{equation}
J_{\rm eff}(n,T) = J n \left[ 1 + \frac{1}{\pi n} \int_0^\pi dk ~ \cos(2k) ~ n_k^F(n,T) \right].
\label{eq:Jeff}
\end{equation}

At zero temperature the integral on the right-hand side of Eq.~\eqref{eq:Jeff} can be evaluated exactly and we obtain
\begin{equation*}
J_{\rm eff}(n,T=0) = J n \left[ 1 - \frac{\sin (2 \pi n)}{2 \pi n} \right].
\end{equation*}
This expression is in agreement with results from exact Bethe-ansatz calculations for the 1D Fermi-Hubbard model. Using this analytical technique the spin-wave dispersion $\epsilon(p)$ was derived in Ref.~\cite{Coll1974} in the large-$U$ limit considered here,
\begin{equation*}
\epsilon(p) = \frac{\pi}{2} J n \left[ 1 - \frac{\sin (2 \pi n)}{2 \pi n} \right] \left| \sin \l  p/ n\r \right|,
\end{equation*}
where $p$ is the momentum of the spin wave. As expected for a spin-wave in squeezed space, the dispersion relation is given by $\epsilon(p) = \pi/2 J_{\rm eff} | \sin p_{\rm sq} |$, where the momentum $p_{\rm sq}$ in squeezed space is related to real space by rescaling all length scales with the density, $p_{\rm sq} = p / n$.

\section{Comparison to theory: exact diagonalization}

In the main part of the paper, we compared our experimental data to theoretical results from the $t-J^*$ Hamiltonian described in Sec.\ref{secSM:SCsep}. To this end we performed exact numerical calculations of thermal states within the $t-J^*$ model. We implemented the Schwinger-boson representation of our model, see Eq.~\eqref{eq:HtJ}, and considered small systems with periodic boundary conditions. In our calculations, the number of holes and the number of spins were fixed, as well as the total magnetization $M$ in $z$-direction. 

We changed the density by changing the number of holes, and calculated the observables $\mathcal{O}(M)$ of interest separately for all possible values of the magnetization. To compare our results directly with the experimental data, we assumed the following distribution function for the total magnetization,
\begin{equation*}
f(M) = \mathcal{N} e^{- \frac{4 M^2}{2 \l N/2\r} },
\end{equation*}
which approximates the experimental magnetization fluctuations (Fig.~S1 C). Here $\mathcal{N}$ is a normalization which ensures that $\sum_{M} f(M) = 1$ and $N$ is the number of fermions in the chain. The final observables are given by $\mathcal{O} = \sum_{M} f(M)  \mathcal{O}(M)$.

To calculate the ground state of our Hamiltonian, we used a standard numerical Lanczos technique. At finite temperatures, $T>0$, the extension of the Lanczos method introduced in Ref.~\cite{Aichhorn2003} was implemented. We sampled over a few thousand random vectors and worked with a dimension between $50$ and $200$ of the Krylof basis, taking system sizes up to $L=14$ sites.

\end{document}